\documentclass[a4paper,twocolumn,11pt,accepted=2021-10-12]{quantumarticle}
\pdfoutput=1
\usepackage[utf8]{inputenc}
\usepackage[english]{babel}
\usepackage[T1]{fontenc}
\usepackage{amsmath}
\usepackage{hyperref}

\usepackage{tikz}
\usepackage{lipsum}

\usepackage{braket}
\usepackage{amssymb}
\usepackage{bm}
\usepackage[numbers,sort&compress]{natbib}

\DeclareMathOperator*{\argmin}{arg\,min}
\renewcommand{\vec}{\bm}

\begin{document}

\title{Simultaneous Perturbation Stochastic Approximation of the Quantum Fisher Information}

\author{Julien Gacon}
\affiliation{IBM Quantum, IBM Research -- Zurich, CH-8803 Rüschlikon, Switzerland
}
\affiliation{Institute of Physics, École Polytechnique Fédérale de Lausanne (EPFL), CH-1015 Lausanne, Switzerland
}
\author{Christa Zoufal}
\affiliation{IBM Quantum, IBM Research -- Zurich, CH-8803 Rüschlikon, Switzerland
}
\affiliation{Institute for Theoretical Physics, ETH Zurich, CH-8092 Zürich, Switzerland
}
\author{Giuseppe Carleo}
\affiliation{Institute of Physics, École Polytechnique Fédérale de Lausanne (EPFL), CH-1015 Lausanne, Switzerland
}
\author{Stefan Woerner}
\email{wor@zurich.ibm.com}
\affiliation{IBM Quantum, IBM Research -- Zurich, CH-8803 Rüschlikon, Switzerland
}

\begin{abstract}
The Quantum Fisher Information matrix (QFIM) is a central metric in promising algorithms, such as Quantum Natural Gradient Descent and Variational Quantum Imaginary Time Evolution.
Computing the full QFIM for a model with $d$ parameters, however, is computationally expensive and generally requires $\mathcal{O}(d^2)$ function evaluations.
To remedy these increasing costs in high-dimensional parameter spaces, we propose using simultaneous perturbation stochastic approximation techniques to approximate the QFIM at a constant cost.
We present the resulting algorithm and successfully apply it to prepare Hamiltonian ground states and train Variational Quantum Boltzmann Machines.
\end{abstract}

\section{Introduction}\label{sec:introduction}

Quantum computing promises potential advances in many fields, such as quantum chemistry and physics \cite{aspuruguzik_simulated_2005, peruzzo_vqe_2014, banuls_lgt_2020}, biology \cite{perdomoortiz_protein_2012, fingerhut_protein_2018, robert_protein_2019}, optimization \cite{farhi_qaoa_2014, gilliam_grover_2019, braine_mixed_2019, gacon_qsbo_2020}, finance \cite{egger_finance_2020}, and machine learning \cite{otterbach_unsupervised_2017, havlicek_supervised_2019, schuld_kernels_2021}.
While fault-tolerant quantum computers are not yet in reach, a computational paradigm particularly suitable for near-term, noisy quantum devices is that of variational quantum algorithms. 
These consist of a feedback loop between a classical and a quantum computer, where the objective function, usually based on a parameterized 
quantum circuit, is evaluated on the quantum computer and a classical counterpart updates the parameters to find their optimal value
\cite{moll_variational_2018}.

In this context, Variational Quantum Imaginary Time Evolution (VarQITE) techniques are particularly promising and received a lot of interest recently \cite{mcardle_varqite_2019, yuan_variational_2018, zoufal_varqbm_2020}. These methods approximate Quantum Imaginary Time Evolution by mapping the quantum state evolution to the evolution of parameters in a parameterized quantum circuit, which serves as an ansatz for the evolved state. 
The interest in these approaches stems from the fact that imaginary time evolution is an integral part of many quantum algorithms and can, for instance, be used to find ground states of given Hamiltonians \cite{mcardle_varqite_2019} or to prepare corresponding Gibbs states \cite{matsui_statistical_1998, masoud_noncommutative_2008, yuan_variational_2018, zoufal_varqbm_2020}. The former is important, e.g., for quantum chemistry or combinatorial optimization, while the latter finds applications, e.g., in the simulation of many-body systems \cite{eisert_manybody_2015}, quantum semi-definite program solvers \cite{brandao_qsdp_2017}, as well as in evaluating and training Quantum Boltzmann Machines (QBMs) \cite{amin_qbm_2018}. 

Another interesting property is that VarQITE is closely related to Quantum Natural Gradient (QNG) Descent \cite{stokes_qng_2020}. 
Unlike standard Gradient Descent, which moves into the steepest direction of the loss function in $\ell_2$ geometry, the QNG considers the steepest direction in the Quantum Information Geometry. This change in geometry has several advantages, such as an invariance under re-parameterization \cite{amari_why_1998} or update steps that are adjusted to the loss sensitivity in each parameter dimension. 
VarQITE coincides with QNG for the special case where the loss function corresponds to the system energy, this is discussed in more detail in Appendix~\ref{app:varqte}.

One significant drawback of VarQITE and QNG is that it requires evaluating the Quantum Fisher Information matrix (QFIM) at every iteration. This operation has a cost scaling quadratically with the number of circuit parameters and is computationally expensive for complex objective function with a large number of variational parameters.
There exist proposed methods to lower the computational cost to linear complexity are to approximate the QFIM by a (block-) diagonal matrix \cite{stokes_qng_2020}, however these cannot properly capture parameter correlations and might not work well for problems where these correlations are strong. 

In this paper, we propose a new approach to approximate VarQITE that only requires a constant number of circuit evaluations per iteration. 
This is achieved by applying ideas originally developed for the Simultaneous Perturbation Stochastic Approximation (SPSA) algorithm \cite{spall_spsa_1998} to approximate the QFIM. A similar approach has previously been explored for the classical Fisher Information matrix in the context of the Expectation-Maximization algorithm \cite{meng_em_2016}.
Our approach is particularly efficient if no precise state evolution is required, as is the case, e.g., for ground state approximation.
However, by allowing additional circuit evaluations, our algorithm is able to approximate the exact path of VarQITE arbitrarily closely.

The remainder of this paper is structured as follows. Sec.~\ref{sec:spsa} reviews first- and second-order SPSA and introduces the required concepts.
Sec.~\ref{sec:nspsa} adapts second-order SPSA to provide stochastic approximations of the QFIM and shows how this can be used to approximate QNG and to train QBMs.
Lastly, we show numerical results for both applications in Sec.~\ref{sec:results} and conclude our paper in Sec.~\ref{sec:conclusion}.

\section{SPSA}\label{sec:spsa}

Minimizing a function's value by selecting optimal input parameters is an ubiquitous problem in computational science, for example, in neural networks or variational quantum algorithms.
A widely used family of methods to find the minimum of the function is gradient descent.
There, starting from an initial guess, the function's parameters are updated iteratively by following the direction of the negative gradient of the function with respect to the parameters. Since the negative gradient points to the direction of steepest descent, the idea is that this update rule will eventually lead to a (local) minimum \cite{cauchy_gd_1847}.

Calculating gradients of a function scales linearly with the number of parameters for both analytic gradients and finite difference approximations. 
As gradient descent techniques require that the gradients must be evaluated at each iteration step, this possibly leads to a computational bottleneck when applied to problems with high-dimensional parameter spaces. 

Simultaneous perturbation methods provide a solution to these linearly increasing computational costs.
Instead of considering each parameter dimension individually, SPSA uses a stochastic approximation for the gradient by simultaneously perturbing all parameters in a random direction. 
This results in an unbiased estimator for the gradient if the random directions are sampled from a suitable distribution, for instance, uniformly from $\{1, -1\}$ for each parameter.
In addition to the computational efficiency, SPSA is also well suited for optimizing noisy objective functions which usually appear in near-term variational quantum algorithms \cite{spall_spsa_1998}.

Let $f: \mathbb{R}^d \rightarrow \mathbb{R}$ be a function with $d$ parameters.
For an initial point $\vec\theta^{(0)} \in \mathbb{R}^d$ and a small learning rate $\eta > 0$, the $k$-th iteration of standard---also called vanilla---gradient descent to minimize $f$ is defined by
\begin{equation}
\label{eq:vanilla_gradient_descent_update}
    \vec\theta^{(k+1)} = \vec\theta^{(k)} - \eta \vec\nabla f(\vec\theta^{(k)}),
\end{equation}
where $\vec\nabla f(\vec\theta) \in \mathbb{R}^d$ denotes the gradient of $f$ with respect to all its parameters.
In contrast, the $k$-th iteration of SPSA first samples a random direction $\vec\Delta^{(k)} \sim \mathcal{U}(\{1, -1\}^d)$ and then approximates the gradient $\vec\nabla f(\vec\theta^{(k)})$ by
\begin{equation}
    \vec\nabla f(\vec\theta^{(k)}) \approx \frac{f(\vec\theta^{(k)} + \epsilon\vec\Delta^{(k)}) - f(\vec\theta^{(k)} - \epsilon\vec\Delta^{(k)})}{2\epsilon} \vec\Delta^{(k)},
\end{equation}
for some small displacement $\epsilon > 0$.
This update uses only two evaluations of $f$, as opposed
to the $\mathcal{O}(d)$ evaluations required for analytic gradients or finite difference approximations.

SPSA can be extended to a second order-method, i.e., to approximate the Hessian in addition to the gradient \cite{spall_2spsa_1997}, and we denote this algorithm as 2-SPSA. 
In second order-methods, the gradient descent update rule is given by
\begin{equation}\label{eq:second_order_gd}
    \vec\theta^{(k+1)} = \vec\theta^{(k)} - \eta H^{-1}(\vec\theta^{(k)}) \vec\nabla f(\vec\theta^{(k)}),
\end{equation}
where $H \in \mathbb{R}^{d \times d}$ is the (approximated) Hessian of $f$.

Instead of computing all $d^2$ entries of $H$ in each iteration, 2-SPSA samples the Hessian using two random directions $\vec\Delta_1^{(k)}$ and $\vec\Delta_2^{(k)}$. The resulting symmetric point-sample is 
\begin{equation}\label{eq:point_estimate}
    \hat{H}^{(k)} = \frac{\delta f}{2\epsilon^2}
                        \frac{\vec\Delta_1^{(k)}\vec\Delta_2^{(k) T} + \vec\Delta_2^{(k)}\vec\Delta_1^{(k) T}}{2},
\end{equation}
where
\begin{equation}\label{eq:delta_f}
    \begin{aligned}
      \delta f &= f(\vec\theta^{(k)} + \epsilon\vec\Delta_1^{(k)} + \epsilon\vec\Delta_2^{(k)}) \\
      &- f(\vec\theta^{(k)} + \epsilon\vec\Delta_1^{(k)}) \\
      &- f(\vec\theta^{(k)} - \epsilon\vec\Delta_1^{(k)} + \epsilon\vec\Delta_2^{(k)}) \\
      &+ f(\vec\theta^{(k)} - \epsilon\vec\Delta_1^{(k)}). 
    \end{aligned}
\end{equation}
The point sample $\hat{H}^{(k)}$ is then combined with all previous samples in an exponentially smoothed estimator
\begin{equation}
    \bar{H}^{(k)} = \frac{k}{k + 1} \bar{H}^{(k - 1)} + \frac{1}{k + 1} \hat{H}^{(k)}.
\end{equation}
To evaluate the gradient, the first-order SPSA technique is used.
In total, this update step uses 6 function evaluations instead of $d^2 + d$ for an analytic second order method, assuming access to the corresponding derivatives.

In Eq.~\eqref{eq:point_estimate}, the Hessian estimate is based on the sampling of two random directions and the resulting point-estimate $\hat{H}^{(k)}$ is an unbiased estimator of the full Hessian.
By re-sampling additional directions and averaging over many point-samples, the stochastic approximation of the Hessian can be systematically improved to arbitrary accuracy \cite{spall_2spsa_1997}. 
Note, however, that the parameter update rule in Eq.~\eqref{eq:second_order_gd} uses the inverse of the smoothed estimator $\bar{H}^{(k)}$ which is not a unbiased estimator of the inverse Hessian anymore. 
While the convergence proofs in Ref.~\cite{spall_2spsa_1997} do not require an unbiased estimator of the inverse Hessian, techniques to remove the bias might improve the convergence, but this is beyond the scope of our work.

Close enough to a minimum, the Hessian of a function is positive semi-definite and our approximation should reflect this.
One possibility for imposing this property is to replace $\bar{H}^{(k)}$ with $\sqrt{\bar{H}^{(k)} \bar{H}^{(k)}}$, whose eigenvalues correspond to the absolute values of the eigenvalues of $\bar{H}^{(k)}$.
Since we also need to ensure invertibility of the Hessian estimate, we further add a small positive regularization constant $\beta > 0$ to the diagonal
and obtain the regularization
\begin{equation}
    \sqrt{\bar{H}^{(k)} \bar{H}^{(k)}} + \beta I,
\end{equation}
where $I \in \mathbb{R}^{d \times d}$ denotes the identity matrix.
To further mitigate instabilities that may arise from a close-to-singular estimate, a blocking condition can be invoked that only accepts an update step $\vec\theta^{(k+1)}$ if the loss at the candidate parameters is smaller than the current loss, plus a tolerance, and otherwise re-samples from the Hessian and gradient.
If the loss function is not evaluated exactly, such as in the case of a sample-based estimation through measurements from a quantum circuit, Ref.~\cite{spall_2spsa_1997} suggests choosing a tolerance that is twice the standard deviation of the loss.

Moreover, the convergence of SPSA and 2-SPSA is guaranteed if a set of conditions on the noise in the loss function evaluation, the differentiability of the loss function, and the meta-parameters $\eta, \epsilon$ and $\vec\Delta$ are satisfied. For details as well as the proof of convergence, we refer to Sec.~3 in Ref.~\cite{spall_2spsa_1997}.

\section{SPSA of the QFIM}\label{sec:nspsa}

In this section, we present the Quantum Natural SPSA (QN-SPSA) algorithm by 
extending 2-SPSA to estimate the QFIM instead of the Hessian of the loss function.
First, we show how QN-SPSA efficiently approximates the QNG algorithm for preparing Hamiltonian ground states.
Then, we leverage this idea to approximate Gibbs state preparation, which we use for the evaluation and training of VarQBMs.
These are all algorithms that rely on accessing the QFIM in every iteration and any algorithm with this reliance can be significantly sped up with our approach.

\subsection{Quantum Natural Gradient}\label{sec:qng}

Consider a parameterized model $p$ depending on $d$ real-valued parameters and a loss function $f$ such that
the loss for parameters $\vec\theta \in \mathbb{R}^d$ is given as $f(p(\vec\theta))$.
Now, the goal is to find the optimal parameters that minimize the loss, given a starting point $\vec\theta^{(0)} \in \mathbb{R}^d$. For convenience we will omit $p$ in the loss expression and abbreviate $f(\vec\theta) = f(p(\vec\theta))$. 

Vanilla gradient descent attempts to minimize the loss by choosing the parameter update step 
proportional to the negative gradient $-\eta\vec\nabla f(\vec\theta)$, with the learning rate $\eta$.
From the geometric perspective, this means selecting the direction of steepest descent in the $\ell_2$ geometry of the loss landscape which induces the smallest possible change in the parameter space. Eq.~\eqref{eq:vanilla_gradient_descent_update} follows, thus, as shown in App.~\ref{app:argmin_formulation}, from the minimization of the following function
\begin{equation}\label{eq:vanilla_argmin}
    \begin{aligned}
   \vec\theta^{(k + 1)} = \argmin_{\vec\theta \in \mathbb{R}^d} \bigg(& \langle \vec\theta - \vec\theta^{(k)}, \vec\nabla f(\vec\theta^{(k)}) \rangle \\
   &+ \frac{1}{2\eta} ||\vec\theta - \vec\theta^{(k)}||_2^2\bigg),
    \end{aligned}
\end{equation}
where $||\cdot||_2$ is the $\ell_2$ norm.
A limitation of vanilla gradient descent is that the learning rate is a global quantity that does not take into account the sensitivity of the model to parameters changes. 
For example, small changes in some subset of parameters might lead to very large changes in model space, whereas large changes of some other parameters be negligible.

An elegant solution to this sensitivity problem is to modify the update step in such a way that the changes with respect to the model $p$ remain under control.
Taking an information geometric approach, this is realized by considering updates in a space that directly reflects the sensitivity of the model. 
To this end, we replace the $\ell_2$ norm $||\cdot||_2$ by
$||\cdot||_{g(\vec\theta)} = \langle\cdot, g(\vec\theta)\cdot\rangle$ where $g(\vec\theta) \in \mathbb{R}^{d \times d}$ denotes the Riemannian metric tensor induced by the model $p(\vec\theta)$. 
Now, the update rule changes to
\begin{equation}
\begin{aligned}
   \vec\theta^{(k + 1)} = \argmin_{\vec\theta \in \mathbb{R}^d}\bigg(& \langle \vec\theta - \vec\theta^{(k)}, \vec\nabla f(\vec\theta^{(k)}) \rangle \\
   & + \frac{1}{2\eta} ||\vec\theta - \vec\theta^{(k)}||_{g(\vec\theta^{(k)})}^2\bigg),
\end{aligned}
\end{equation}
which---as shown in App.~\ref{app:argmin_formulation}---results in the Natural Gradient Descent formula \cite{amari_why_1998}
\begin{equation}\label{eq:natural_gradient}
    \vec\theta^{(k+1)} = \vec\theta^{(k)} - \eta g^{-1}(\vec\theta^{(k)}) \vec\nabla f(\vec\theta^{(k)}).
\end{equation}
See also Refs.~\cite{stokes_qng_2020, yao_opticalqng_2021} for more detailed discussions on the derivation of the QNG.

We now consider the case where $p$ is a parameterized quantum circuit.
Let $\ket{\psi(\vec\theta)}$ describe a parameterized pure quantum state in a Hilbert space on $n$ qubits for $d$ classical parameters $\vec\theta \in \mathbb{R}^d$. 
Then, the metric tensor $g(\vec\theta) \in \mathbb{R}^{d \times d}$ is the Fubini-Study metric tensor with elements \cite{stokes_qng_2020} 
\begin{equation}\label{eq:qfi_direct}
    \begin{aligned}
    g_{ij}(\vec\theta) = \mathrm{Re}\bigg\{
        \left\langle \frac{\partial \psi}{\partial {\theta_i}} \bigg\vert \frac{\partial \psi}{\partial \theta_j}\right\rangle 
        - \left\langle \frac{\partial \psi}{\partial \theta_i} \bigg\vert \psi \right\rangle
        \bigg\langle \psi \bigg\vert \frac{\partial \psi}{\partial \theta_j} \bigg\rangle
    \bigg\},
    \end{aligned}
\end{equation}
where $\theta_i$ ($\theta_j$) denotes the $i$-th ($j$-th) element of the parameter vector $\vec\theta$ and the quantum state and its derivatives are evaluated at $\vec\theta$.
The required expectation values can be computed by using a linear combination of unitaries or by parameter shift techniques \cite{schuld_evaluating_2019}. Note that evaluating the Fubini-Study metric tensor $g$ allows us to directly compute the QFIM since they are equivalent up to a constant factor; the QFIM equals $4g$ \cite{meyer_fisher_2021}.

Computing $g$ in general requires evaluating $\mathcal{O}(d^2)$ expectation values. By using the 2-SPSA algorithm, we can replace $g(\vec\theta^{(k)})$ by a stochastic approximation $\bar g^{(k)}$, requiring only the evaluation of four expectation values, i.e., constant and independent of $d$.
To exploit 2-SPSA, we use a different representation of the metric tensor $g$ than in Eq.~\eqref{eq:qfi_direct}, namely, the Hessian of the Fubini-Study metric \cite{stokes_qng_2020, mari_estimating_2020}
\begin{equation}\label{eq:qfi_hessian}
   g_{ij}(\vec\theta) = -\frac{1}{2}\frac{\partial}{\partial \theta_i} \frac{\partial}{\partial \theta_j} \vert \langle\psi(\vec\theta^\prime) | \psi(\vec\theta)\rangle\vert^2 \bigg\vert_{\vec\theta^\prime = \vec\theta}.
\end{equation}
See Appendix~\ref{app:qfi_reformulation} for the equivalence to the previous representation.
We generalize 2-SPSA for the Hessian of a metric instead of a function by applying perturbations only to the second argument of the metric and keeping the first argument fixed. 
Concretely, Eqs.~\eqref{eq:point_estimate} and~\eqref{eq:delta_f} change to
\begin{equation}
    \hat{g}^{(k)} = -\frac{1}{2} \frac{\delta F}{2\epsilon^2}
                    \frac{\vec\Delta_1^{(k)}\vec\Delta_2^{(k) T} + \vec\Delta_2^{(k)}\vec\Delta_1^{(k) T}}{2},
\end{equation}
where
\begin{equation}
    \begin{aligned}
    \delta F &= F(\vec\theta^{(k)}, \vec\theta^{(k)} + \epsilon\vec\Delta_1^{(k)} + \epsilon\vec\Delta_2^{(k)}) \\
      &- F(\vec\theta^{(k)}, \vec\theta^{(k)} + \epsilon\vec\Delta_1^{(k)}) \\
      &- F(\vec\theta^{(k)}, \vec\theta^{(k)} - \epsilon\vec\Delta_1^{(k)} + \epsilon\vec\Delta_2^{(k)}) \\
      &+ F(\vec\theta^{(k)}, \vec\theta^{(k)} - \epsilon\vec\Delta_1^{(k)}),
    \end{aligned}
\end{equation}
and $F(\vec\theta, \vec\theta^\prime) = \vert \langle\psi(\vec\theta) | \psi(\vec\theta^\prime)\rangle\vert^2$. The 
smoothing of the point-estimates $\hat g^{(k)}$ into $\bar g^{(k)}$ and 
the technique to ensure the estimate is positive semi-definite remains the same as in the previous section.

Evaluating the Fubini-Study metric requires calculation of the absolute value of the overlap of $\ket{\psi(\vec\theta)}$ with parameter values $\vec\theta$ and slightly shifted parameters $\vec\theta + \epsilon\vec\Delta$. 
The overlap of two quantum states can for instance be estimated using the swap test \cite{buhrman_swaptest_2001}, where both states are prepared in separate qubit registers.
Another option, if the states are given by $\ket{\psi(\vec\theta)} = U(\vec\theta)\ket{0}$ for a parameterized unitary $U$, and we only need the absolute value of the overlap, is to prepare $U^\dagger(\vec\theta + \epsilon\vec\Delta) U(\vec\theta)\ket{0}$ and estimate the probability of measuring $\ket{0}$, which is equal to $|\langle \psi(\vec\theta)| \psi(\vec\theta + \epsilon\vec\Delta) \rangle |^2$.
If our state has $n$ qubits and the circuit corresponding to $U$ has depth $m$, the swap test requires a circuit width of 
$2n$, but only leads to a depth of $m + \mathcal{O}(1)$ \cite{cincio_learning_2018}.
In contrast, the compute-uncompute method \cite{havlicek_supervised_2019} uses circuits of width $n$, but instead needs twice the depth, $2m$.
To avoid doubling both circuit with and circuit depth, the overlap can also be estimated via randomized measurements of two independent state preparations \cite{elben_overlap_2019}, however this technique requires an exponential number of measurements.
Depending on the complexity of the unitary and the structure of the available hardware, either method can be advantageous.

\subsection{Quantum Boltzmann Machines}\label{sec:qbm}

QBMs are energy-based machine learning models that encode information in the parameters $\vec\omega$ of a parameterized $n$-qubit Hamiltonian $\hat{\mathcal{H}}_{\vec\omega}$ \cite{amin_qbm_2018}. This Hamiltonian defines a Gibbs state
\begin{equation}\label{eq:gibbs_state}
	\rho^{\text{Gibbs}}(\hat{\mathcal{H}}_{\vec\omega}) = \frac{e^{-\hat{\mathcal{H}}_{\vec\omega}/\left(\text{k}_{\text{B}}\text{T}\right)}}{Z},
\end{equation}
with the Boltzmann constant $\text{k}_{\text{B}}$, system temperature $T$ and partition function $Z=\text{Tr}[e^{-\hat{\mathcal{H}}_{\vec\omega}/(\text{k}_{\text{B}} T)}]$.
Depending on the construction of $\hat{\mathcal{H}}_{\vec\omega}$ and the choice of the loss function, QBMs can be used for various machine learning tasks, such as generative or discriminative learning \cite{zoufal_varqbm_2020}.
Throughout the training, the Gibbs state $\rho^\text{Gibbs}(\hat{\mathcal{H}}_{\vec\omega})$ is repeatedly prepared and measured for different parameter values $\vec\omega$.
The obtained samples are then used to evaluate the loss function.

Preparing quantum thermal states, such as Gibbs states, is difficult and several techniques have been proposed to solve this task \cite{Temme2011QuantumMS, YungQuantumMetropolis12, PoulinThermalQGibbs09, MottaQITE20, brandaoFiniteCorrLengthEfficientPrep19, BrandaoGibbsSampler16, VarThermofieldDSJingxiang19, WiebeVariationalGibbs2020}.
In the following, we consider approximate construction of the Gibbs state using VarQITE, which follows the time evolution of a maximally mixed state under $\hat{\mathcal{H}}_{\vec\omega}$ for the time $(2\text{k}_{\text{B}}\text{T})^{-1}$. 

At first, the initial maximally mixed state on $n$ qubits is constructed using $n$ additional environmental qubits. Each of the $n$ qubits encoding the Gibbs state is assigned to one environmental qubit and each qubit pair is prepared in the Bell state $(\ket{00} + \ket{11}) / \sqrt{2}$. If the environmental qubits are now traced out, the remaining $n$ qubits are in a maximally mixed state. The Hamiltonian is adjusted to the extended system by acting trivially on the environmental qubits
\begin{equation*}
    \hat{\mathcal{H}}_{\vec\omega} \rightarrow \hat{\mathcal{H}}_{\vec\omega} \otimes \mathbb{I}^{\otimes n},
\end{equation*} 
where $\mathbb{I}$ denotes the identity operator on a single qubit.
In the variational approach, the state of the $2n$ qubits is represented by a parameterized quantum circuit with parameters $\vec\theta$. For VarQBMs, the initial parameter values $\vec\theta^{(0)}$ must be chosen such that each qubit pair is in a Bell state \cite{zoufal_varqbm_2020}.

With the correct initial state prepared, we can now apply VarQITE. 
The update rule for the circuit parameters are governed by McLachlan's variational principle \cite{mclachlan_variational_1964}
\begin{equation}
\label{eq:McLachlan}
    g_{ij}(\vec\theta^{(t)})\frac{\partial\theta_j^{(t)}}{\partial t} = -\text{Re}\left\{ \left\langle\frac{\partial\psi(\vec\theta^{(t)})}{\partial \theta_i}\right\vert \hat{\mathcal{H}}_{\vec\omega} \ket{\psi(\vec\theta^{(t)})} \right\},
\end{equation}
where $g$ is the Fubini-Study metric from Eq.~\eqref{eq:qfi_direct}. We obtain the time-evolution of the parameters by integrating with an arbitrary ODE solver, such as the explicit Euler scheme
\begin{equation}
    \vec\theta^{(t + \delta\tau)} = \vec\theta^{(t)} + \delta \tau \frac{\partial\vec\theta^{(t)}}{\partial t},
\end{equation}
where $\delta \tau$ is $(2\mathrm{k}_\mathrm{B} T)^{-1}$ divided by the number of time steps.

Now, we can apply the same idea as before and replace $g(\vec\theta^{(t)})$ in the linear system of equations in Eq.~\eqref{eq:McLachlan} with the approximation $\bar{g}^{(t)}$ obtained with QN-SPSA and hence significantly reduce the costs associated with Gibbs state preparation, while sacrificing some accuracy.
Note that for VarQITE we attempt to track the exact evolution more closely than for the QNG and therefore it 
is important to average over multiple point samples $\hat{g}^{(t)}$ per time step.

\section{Numerical Results}\label{sec:results}

In this section, we apply the introduced technique to different problem instances. First, we analyze how QN-SPSA performs compared with QNG for ground state approximation, and, second, we show how VarQBMs perform when the Gibbs states are prepared with VarQITE when the QFIM is approximated using 2-SPSA.

\subsection{Diagonal Hamiltonians}\label{sec:numerics_qng}

\begin{figure*}
    \centering
    \begin{tikzpicture}
\node[inner sep=0pt, anchor=north west] at (-10.4, -0.5) {\includegraphics[width=0.3\textwidth]{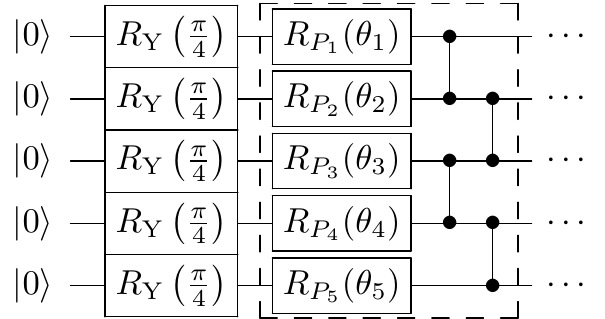}};
\node[anchor=north west] at (-5,0) {\includegraphics[width=0.34\textwidth]{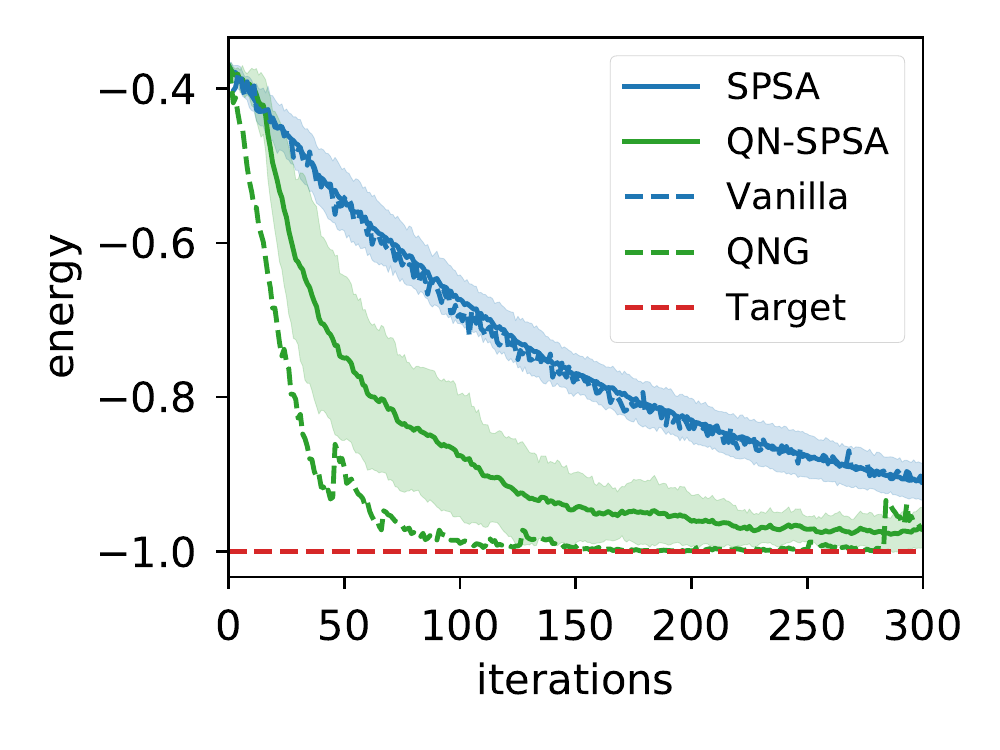}};
\node[anchor=north west] at (1,0) {\includegraphics[width=0.34\textwidth]{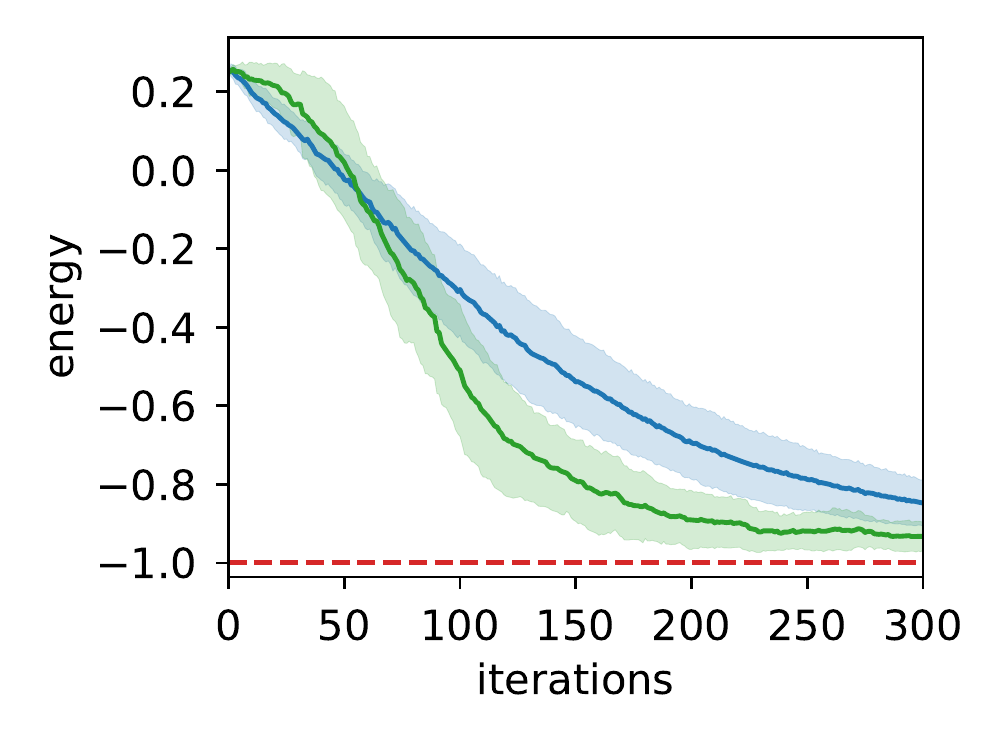}};
\node at (-10, 0) {(a)};
\node at (-4.6, 0) {(b)};
\node at (1.4, 0) {(c)};
\end{tikzpicture}
    \caption{Investigation of the loss for the Pauli two-design circuit with the local observable $\hat{\mathcal{H}} = Z_1 Z_2$.
    (a) The circuit for five qubits, where in each layer the dashed box is repeated. At the end we add a final rotation layer, not shown here. For each rotation gate $R_{P_i}$ the rotation axis is chosen uniformly at random, i.e., $P_i \sim \mathcal{U}(\{X, Y, Z\})$.
    (b) The loss for 11 qubits with the observable $Z_5 Z_6$ and 3 layer repetitions for vanilla gradient descent, QNG and the respective SPSA variants.
    (c) The same problem scaled to 22 qubits with the observable $Z_{11} Z_{12}$ and 5 layer repetitions. The analytic optimizers are not shown since they are computationally too costly to evaluate for 132 parameters.}
    \label{fig:pauli2design}
\end{figure*}
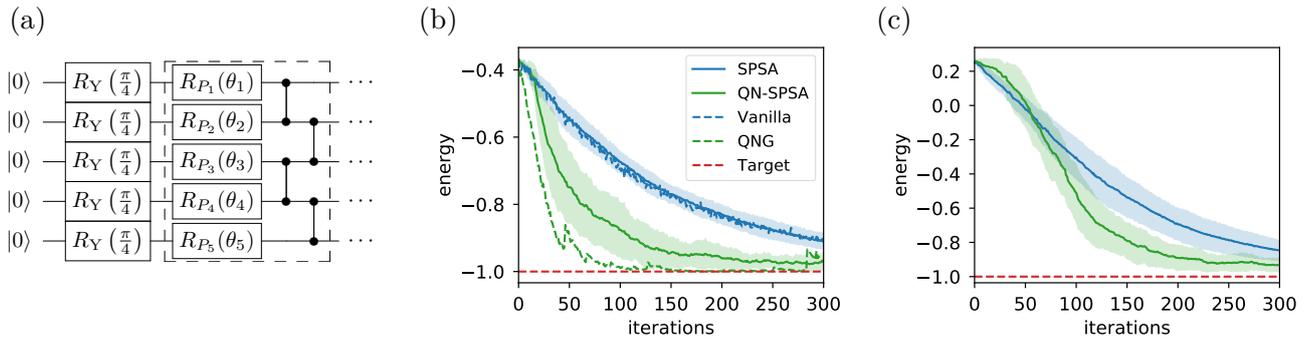

We compare the performance of the following optimization routines on problems from the related literature to illustrate the speed of convergence, as well as the robustness of convergence with respect to the initial choice of parameters: vanilla gradient descent, SPSA, QNG, and QN-SPSA.

\subsubsection{Speed of convergence}
 
To investigate the speed of convergence, we compare the value of the loss function against the number of 
iterative parameter updates. As in Ref.~\cite{stokes_qng_2020} we use 
a Pauli two-design circuit as the parameterized quantum circuit.
The circuit consists of an initial layer of $R_\mathrm{Y}(\pi/4)$ gates followed by alternating rotation and 
entanglement layers. The rotation layers apply uniformly at randomly selected $R_\mathrm{X}, R_\mathrm{Y}$ or $R_\mathrm{Z}$ gates and
the entanglement layers apply controlled-$Z$ gates between nearest neighbors.
An example of the circuit for 5 qubits is schematically shown in Fig.~\ref{fig:pauli2design}(a).
The loss function is the expectation with respect to a local $Z_{i} Z_{i+1}$ observable placed in the 
middle of the circuit and the layers in the circuit are repeated sufficiently often such that the light cone of the two measurements involves all qubits. For instance, for 11 qubits we set $i = 5$ and repeat the layers 3 times.

In this benchmark, we use 11 and 22 qubits with 3 and 5 layer repetitions leading to 44 and 132 parameters respectively.
The learning rate for all optimizers is chosen to be $\eta=10^{-2}$ and the displacement for the finite different approximation in the SPSA methods is $\epsilon=10^{-2}$.
The regularization constant for QN-SPSA is set to $\beta=10^{-3}$.
The analytic optimizers are only run once, while for both SPSA techniques we show the mean and standard deviation of 25 independent runs from the same initial point.
The circuits are implemented in Qiskit \cite{qiskit_2019} and executed using the built-in simulator with 8192 shots.

Fig.~\ref{fig:pauli2design}(b) shows the loss per iteration for each method for 11 qubits.
As previously presented in Ref.~\cite{stokes_qng_2020}, the analytic QNG converges faster than vanilla gradient descent.
The spikes in the QNG loss are due to numerical instabilities in the inversion of the Fubini-Study metric tensor and can be avoided using a regularization.
The mean of the standard SPSA algorithm coincides almost exactly with the vanilla gradient descent,
which is the expected behavior since SPSA is an unbiased estimator of the gradient.
We further observe that QN-SPSA outperforms SPSA and vanilla gradient descent and approaches the loss achieved by the analytic QNG, although with a larger variance than standard SPSA. 
The mean of the QN-SPSA loss is close to the QNG loss, but it cannot reach it due to the regularization constant $\beta > 0$ that we add for numerical stability. With this regularization constant, we can interpolate
between the natural gradient ($\beta = 0$) and vanilla gradient ($\beta \gg 0$), see Appendix~\ref{app:qfi_regularization} a more detailed investigation.

In Fig.~\ref{fig:pauli2design}(c), we repeat the experiment for 22 qubits. With 132 parameters, this example already manifests the advantage of SPSA-based optimizers over analytic gradients: While QNG requires the execution of approximately 2.6 million circuits, QN-SPSA needs only 2100 circuits. Due to the large computational cost, the analytic gradients are not presented in the 22 qubit case.

In Appendix~\ref{app:maxcut}, we compare the convergence of the different optimization schemes with respect to the number of function evaluations and discuss the efficiency and true costs of the different optimizers in more detail.

\subsubsection{Region of convergence}

The advantage of natural gradients is not just a faster convergence, which---for problems with a simple loss landscape---might also be achieved with vanilla gradient descent or SPSA if the learning rate is carefully calibrated.
But, since QNG (or VarQITE) approximates imaginary time evolution, we have the guarantee that QNG always converges to the ground state if the initial state has a non-zero overlap with it and if a sufficiently powerful ansatz and small stepsize are chosen \cite{mcardle_varqite_2019}. 
Even with an ansatz that cannot follow the imaginary time evolution exactly, QNG and QN-SPSA can have superior convergence properties to vanilla gradient descent and SPSA.

\begin{figure*}
    \centering
    \begin{tikzpicture}
        \node[anchor=north west] at (0, -3.3) {\includegraphics[width=0.25\textwidth]{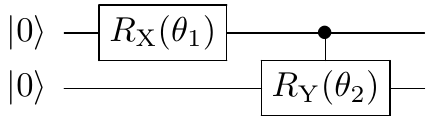}};
        \node at (0.4, -3) {(a)};
        \node[anchor=north west] at (5, 0) {\includegraphics[width=0.74\textwidth]{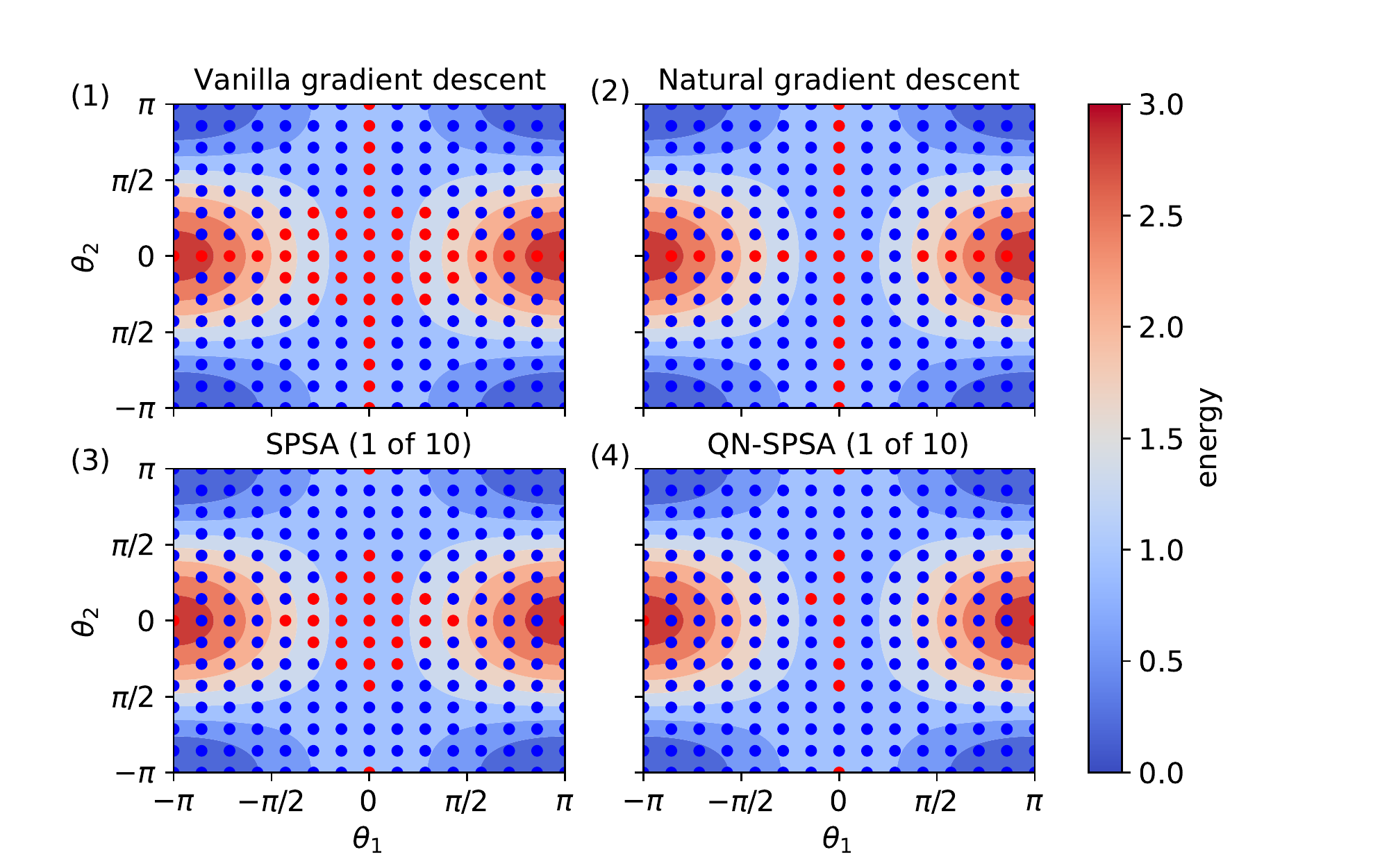}};
        \node at (5.2, -0.5) {(b)};
    \end{tikzpicture}
    \caption{Convergence tests for a simple loss function. (a) The parameterized quantum circuit used as ansatz. (b) A comparison of the performance of (1) vanilla gradient descent, (2) QNG, (3) SPSA and (4) QN-SPSA. Each dot marks an initial point, which is marked blue if the method converged and red otherwise. For SPSA and QN-SPSA we repeat the optimization 10 times and consider the point as converged if at least one out of the 10 runs converged to the global optimum.
    The global phase parameter $\theta_0$ is not shown since it does not affect the value of the loss function.}
    \label{fig:convergence_region}
\end{figure*}

To illustrate this, we use the same problem as in Ref.~\cite{mcardle_varqite_2019} with the ansatz
\begin{equation*}
    \begin{aligned}
    \ket{\psi(\vec\theta)} = &e^{i\theta_0} (\ket{0}\bra{0} \otimes \mathbb{I} + \ket{1}\bra{1} \otimes R_\mathrm{Y}(\theta_2)) \\ &(R_\mathrm{X}(\theta_1) \otimes \mathbb{I}) \ket{00},
    \end{aligned}
\end{equation*}
prepared by the circuit in Fig.~\ref{fig:convergence_region}(a) and try to minimize the energy with 
respect to the Hamiltonian
\begin{equation*}
    \hat{\mathcal{H}} = \begin{pmatrix}
    1 & 0 & 0 & 0 \\
    0 & 2 & 0 & 0 \\
    0 & 0 & 3 & 0 \\
    0 & 0 & 0 & 0
    \end{pmatrix}.
\end{equation*}
A variational global phase is added to account for phase differences between the target state and the ansatz, which does not impact the expectation value but can lead to incorrect gradients \cite{yuan_variational_2018}.
We choose different initial points in the same loss landscape and test if vanilla gradient, natural gradient, and the SPSA variants converge to the optimal solution.

In this example, we choose an equidistant grid of $15 \times 15$ points in $[-\pi, \pi]^2$ for the initial values of $\theta_1$ and $\theta_2$. 
The initial global phase is set to zero, $\theta_0 = 0$. 
As in Ref.~\cite{mcardle_varqite_2019} we use constant learning rates of $\eta=0.886$ for vanilla gradient descent and $\eta=0.225$ for QNG and all methods do 200 iterations. 
We consider an optimization run as converged if the final absolute error is below $10^{-4}$.
For the SPSA methods, we execute 10 optimization runs for each initial point and label the point converged if at least one out of the 10 runs converged. The analytic methods are deterministic and only run once. 
Standard SPSA and QN-SPSA use the same learning rates as the 
corresponding analytic versions, i.e., 0.886 and 0.225, respectively.

The results are shown in Fig.~\ref{fig:convergence_region}(b).
The vanilla gradient descent and QNG reproduce the results from Ref.~\cite{mcardle_varqite_2019}. QNG converges from all sampled points except when one of the initial angles is exactly 0, where at least one gradient component vanishes and the parameter update cannot move towards one of the minima in the corners of the plot.
Vanilla gradient additionally fails to converge in a diamond-shaped region around the saddle point $(0, 0)$.
The regions of convergence for the SPSA methods are similar to the analytic variants, however, they do not suffer as much from vanishing gradient components. Due to the random selection of the direction of the gradient, the stochastic methods have a chance to move to a region where both gradient components are nonzero.
To conclude, QN-SPSA outperforms all other methods and impressively converges for the most initial points.

\subsection{Molecular Hamiltonian}\label{sec:lih}

In this section we use QN-SPSA to approximate the ground state of the lithium-hydride (LiH) molecule at a bond distance of 2.5\AA{} on the \textit{ibmq\_montreal} device, which is one of the IBM Quantum Falcon processors, using Qiskit Runtime \cite{qiskit_runtime_2021}. 

\begin{figure*}
    \centering
    \begin{tikzpicture}
        \node[anchor=north west] at (0, 0) {\includegraphics[width=0.55\textwidth]{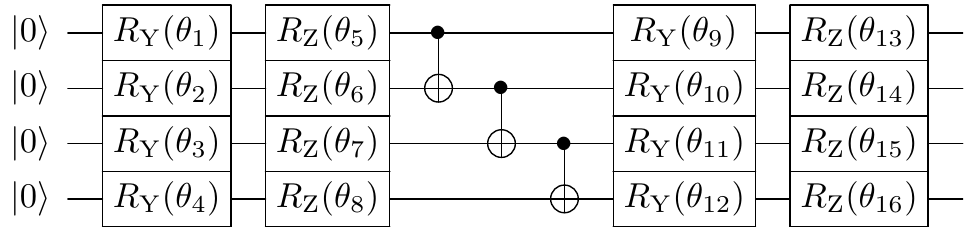}};
        \node at (0.4, 0.3) {(a)};
        \node[anchor=north west] at (11, 1) {\includegraphics[width=0.34\textwidth]{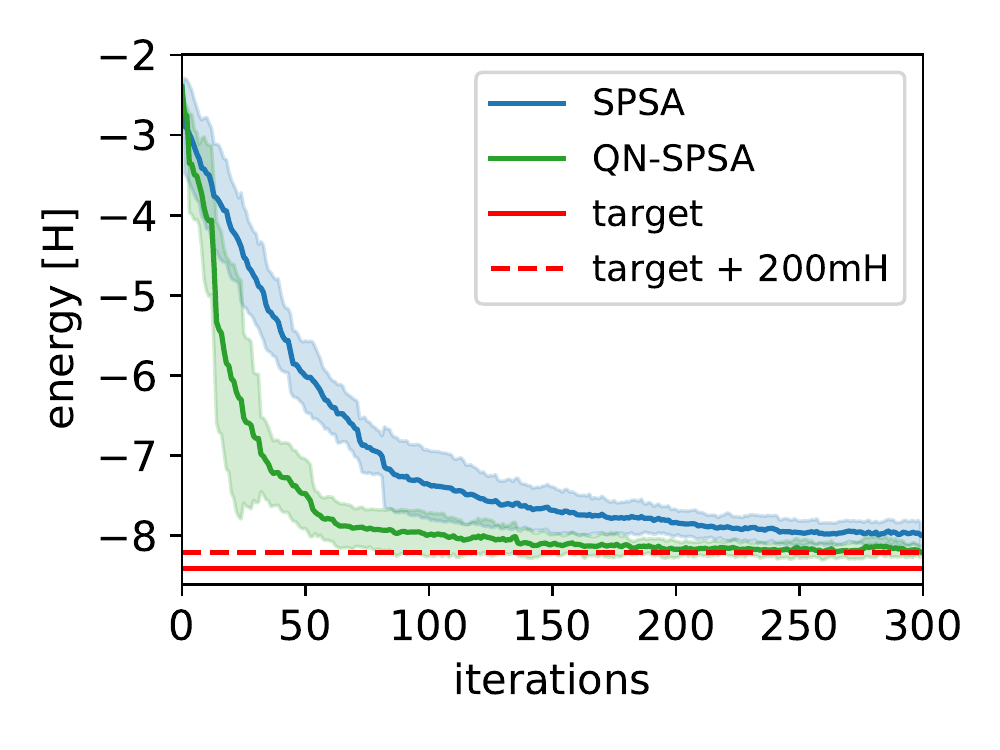}};
        \node at (11, 1.3) {(b)};
    \end{tikzpicture}
    \caption{Ground state calculation of LiH at bond distance of 2.5\AA{}. 
            (a) The hardware-efficient ansatz wave function with alternating rotation and entanglement layers.
            (b) The evolution of the energy for QN-SPSA and SPSA in Hartree. The reference energy is calculated classically using the dense matrix representation of the qubit Hamiltonian.
            } 
    \label{fig:lih}
\end{figure*}

In order to extract the one and two body integrals for the molecular system we perform a Restricted Hartree Fock calculation using PySCF. For the description of the system we use the STO3G basis set that results in 6 molecular orbitals. We restrict the active space of the system to 3 molecular orbitals and we use the parity mapping as fermion-to-qubit transformation \cite{bravyi_tapering_2017}, leading to a system of 6 qubits. The intrinsic property of parity mapping \cite{bravyi_tapering_2017} allows us to reduce the qubit requirements by another 2 qubits and we eventually obtain a system of 4 qubits that we simulate. As trial wave function we select a hardware-efficient ansatz, similar to Ref.~\cite{kandala_hardware_2017}, that consists of alternating single-qubit rotation layers and two-qubit nearest-neighbour entanglement layers. This entanglement structure is naturally compatible with our device's heavy-hex coupling map such that we can map the circuit to our hardware with no additional SWAP gates. The circuit schematics are shown in Fig.~\ref{fig:lih}(a).

In Fig.~\ref{fig:lih}(b) we present the convergence of the QN-SPSA and SPSA optimization using a learning rate of $\eta=10^{-2}$, a perturbation of $\epsilon=10^{-1}$ and a regularization of $\beta=10^{-3}$. The perturbation is larger than in the noiseless simulations, which is a less accurate approximation but much more stable with respect to local fluctuations induced by the noisy loss function evaluations. 
To start with a good estimate for the Fubini-Study metric tensor, the first two iterations of QN-SPSA average 100 point-samples into a single estimate and then average over 2 for the rest of the optimization. 
For both optimizers we start at the same random initial point and show average and standard deviation of 5 experiments with 300 iterations each. We observe that QN-SPSA not only converges faster than SPSA but also reaches a lower final energy. 
Due to hardware noise, the final energy achieved deviates from the exact energy by approximately 200mH, which can likely be overcome using error mitigation techniques such as Richardson Extrapolation \cite{kandala_extending_2019}. 

After grouping all commuting Pauli terms we need to measure in 25 different bases to evaluate the expectation value of the Hamiltonian, where we average each measurement over 1024 shots. Since the QFIM only depends on the ansatz and is independent of the system's Hamiltonian, each stochastic sample of the QFIM still only requires 4 circuit evaluations. Thus, the more complex the problem Hamiltonian gets the smaller the overhead of computing the QN-SPSA update becomes compared to SPSA.

\begin{figure*}
    \centering
        \centering
    \begin{tikzpicture}
        \node[anchor=north west] at (0, 0) {\includegraphics[width=0.34\textwidth]{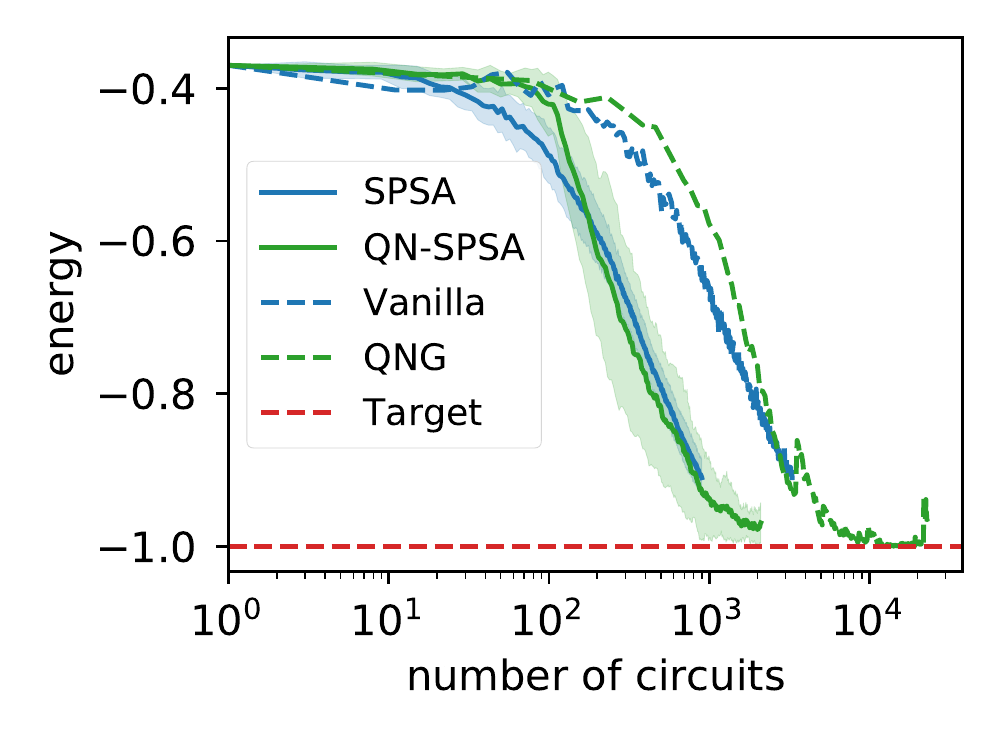}};
        \node at (0.4, 0.2) {(a)};
        \node[anchor=north west] at (8, 0) {\includegraphics[width=0.34\textwidth]{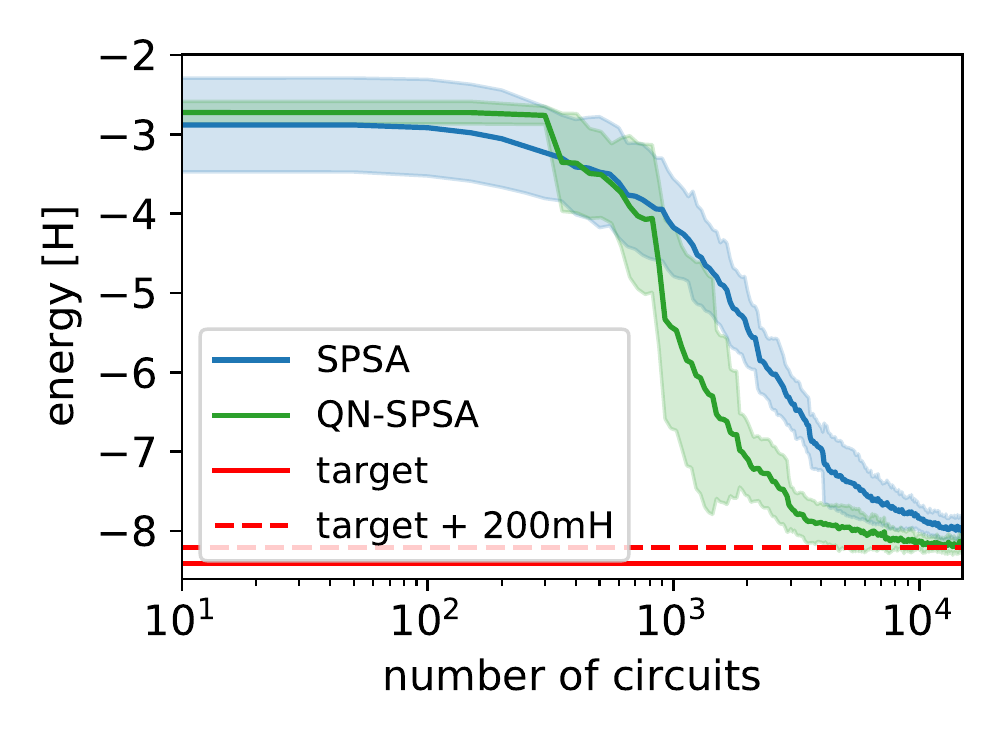}};
        \node at (8, 0.2) {(b)};
    \end{tikzpicture}
    \caption{Convergence comparison with respect to the number of evaluated circuits.
            (a) The Pauli two-design example on 11 qubits.
            (b) The LiH experiment.
            } 
    \label{fig:nfevs}
\end{figure*}

To assess the real computational cost on the quantum device, we change the perspective to see how fast each method converges with respect to the number of function evaluations instead of iterations. One function evaluation corresponds here to the execution of a single circuit.
For the Pauli two-design, see Fig.~\ref{fig:nfevs}(a), we observe that the analytic methods require about one, respectively two, orders of magnitude more evaluations than the SPSA-based techniques. Since QN-SPSA requires 7 function evaluations per step and SPSA only 3, both methods perform similarly for this diagonal Hamiltonian even though QN-SPSA takes less iterations to converge. 
For a non-diagonal Hamiltonian however, such as for LiH where we have to measure in multiple bases, QN-SPSA adds negligible additional costs and significantly speeds up the converge as shown in Fig.~\ref{fig:nfevs}(b).
In Appendix~\ref{app:maxcut} we provide another example where QN-SPSA converges much faster.
If the loss function parameters are much more sensitive in certain dimensions than others, the Natural Gradient information can modulate the learning rate accordingly, whereas vanilla gradients are forced to choose a learning rate small enough to avoid overshooting in any parameter dimension.

\subsection{Quantum Boltzmann Machines}\label{sec:numerics_qbm}

In the following, we show that QN-SPSA enables the realization of an approximate VarQBM implementation, where the computational complexity is reduced compared to standard VarQITE-based Gibbs state preparation. We choose to apply the suggested method to a generative learning example, investigated in Ref.~\cite{zoufal_varqbm_2020}, where the aim is to prepare a Gibbs state whose sampling probabilities correspond to a given target probability density function.

The learning task in this example is to reproduce the sampling statistics of the Bell state
$(\ket{00} + \ket{11})/\sqrt{2}$. Given a parameterized Hamiltonian $\hat{\mathcal{H}}_{\vec\omega}$,
we aim to find a set of parameters, $\vec\omega$, such that the sampling probabilities of the corresponding Gibbs state 
$\rho^\text{Gibbs}(\hat{\mathcal{H}}_{\vec\omega})$ are close to the target sampling probabilities 
\begin{equation}
    p^\text{Bell} = (0.5, 0, 0, 0.5),
\end{equation} 
measured in the computational basis.
The distance to the target distribution is assessed using the relative entropy, which is a measure that describes the distance between two probability distributions. Minimizing the distance between two distributions with respect to the relative entropy is equivalent to minimizing the cross-entropy of the respective probability distributions
\begin{equation*}
    \ell(\vec\omega) = -\sum_{x=0}^{3} p^\text{Bell}_x \log\left(p^\text{Gibbs}_x(\hat{\mathcal{H}}_{\vec\omega})\right),
\end{equation*}
where $x$ corresponds to the computational basis states and
\begin{equation}
    p^\text{Gibbs}_x(\hat{\mathcal{H}}_{\vec\omega}) = \mathrm{Tr}\left[\rho^\text{Gibbs}(\hat{\mathcal{H}}_{\vec\omega})\ket{x}\bra{x} \right].
\end{equation}

\begin{figure*}
    \centering
    \begin{tikzpicture}
        \node[anchor=north west] at (-3, 0) {\includegraphics[width=0.6\textwidth]{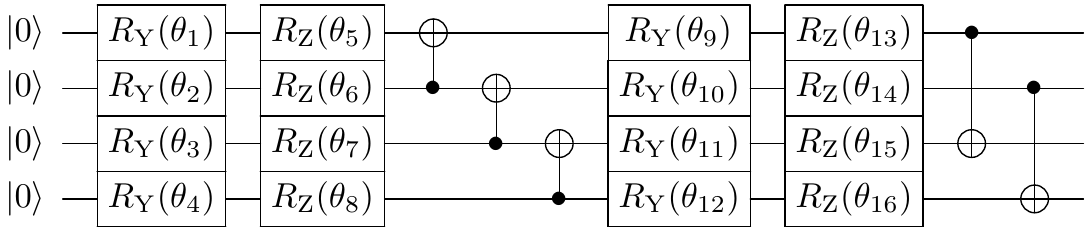}};
        \node[anchor=north west] at (-4, -2.6) {\includegraphics[width=0.32\textwidth]{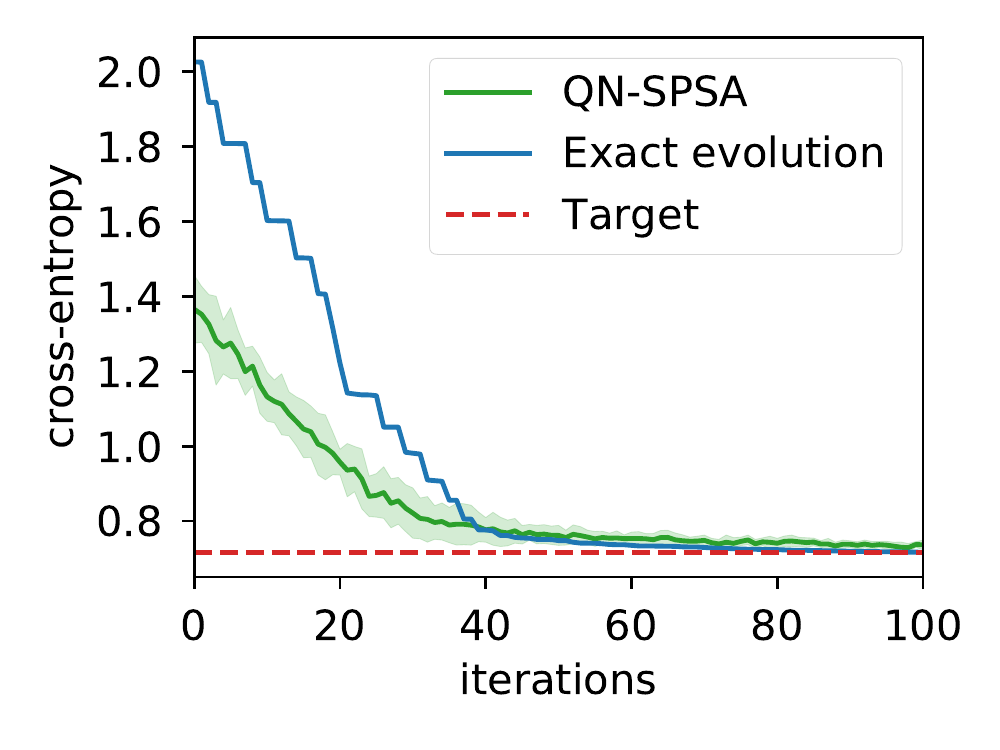}};
        \node[anchor=north west] at (3, -2.6) {\includegraphics[width=0.33\textwidth]{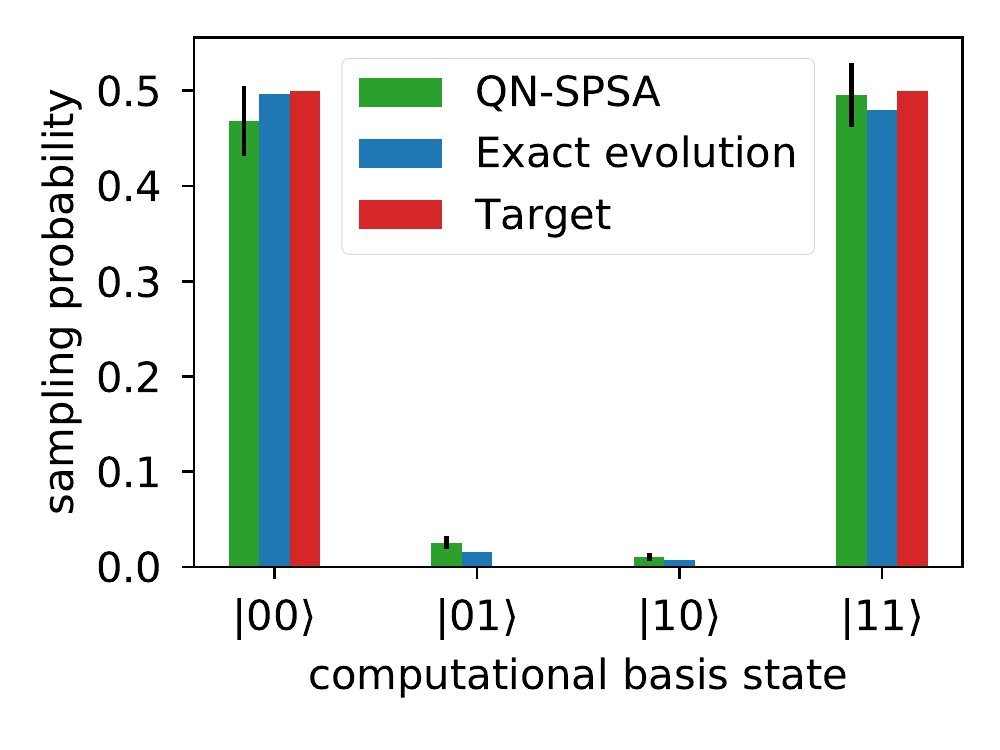}};
        \node at (-2.7, 0.2) {(a)};
        \node at (-3.9, -2.7) {(b)};
        \node at (3.1, -2.7) {(c)};
    \end{tikzpicture}
    \caption{Generative learning with VarQBMs. (a) The parameterized circuit encoding the Gibbs state. (b) The mean and standard deviation of the QN-SPSA training loss and loss of the exact evolution via matrix exponentiation. The target loss is the final value the training with exact evolution. (c) The final probabilities of the trained Gibbs states. For each of the 10 optimization runs we prepare the 
    final state 10 times to approximate the standard deviation on the final sampling statistics.}
    \label{fig:generative_learning}
\end{figure*}
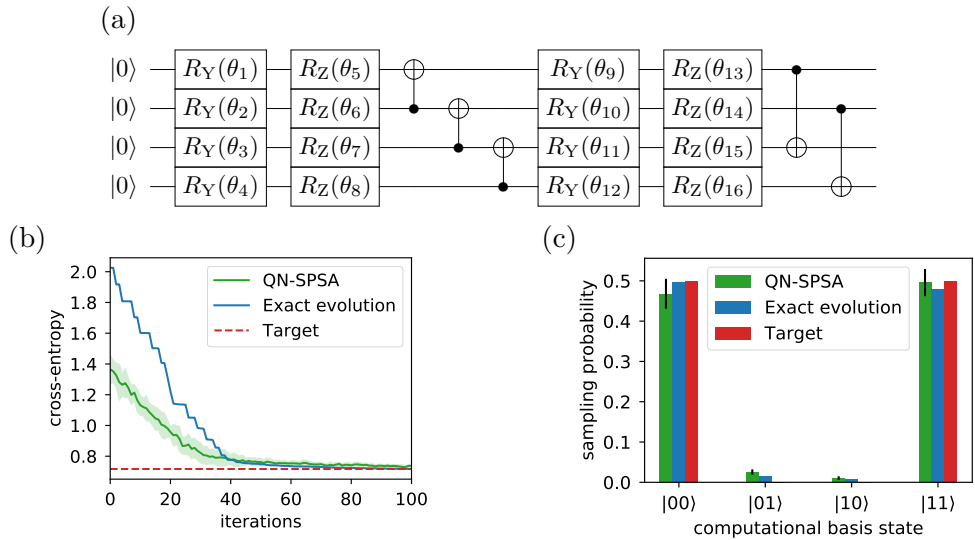

This example uses the parameterized Hamiltonian
\begin{equation*}
    \hat{\mathcal{H}}_{\vec\omega} = \omega_1 Z_1 Z_2 + \omega_2 Z_1 + \omega_3 Z_2.
\end{equation*}
The system temperature is set to $\mathrm{k}_\mathrm{B} T = 1$ which results in the evolution time $(2\mathrm{k}_\mathrm{B}T)^{-1} = 0.5$. Furthermore, the approximate QN-SPSA Gibbs state preparation uses forward Euler with 10 equidistant time steps and the ansatz circuit shown in Fig.~\ref{fig:generative_learning}(a). 
To start the evolution in a maximally mixed state, the initial parameters for the circuit are
\begin{equation*}
    \forall i \in \{1, \dots, d\}: \theta^{(0)}_i = 
    \begin{cases}
        \frac{\pi}{2}, \text{ if } i \in \{9, 10\}, \\
        0, \text{ otherwise.}
    \end{cases}
\end{equation*}
The initial parameters for the Hamiltonian $\vec\omega^{(0)}$ are chosen uniformly at random from $[-2, 2]$.
We optimize the Hamiltonian parameters $\vec\omega$ with 100 iterations of SPSA with a learning rate and perturbation of 0.1.

For QN-SPSA, we choose a perturbation of $\epsilon = 10^{-2}$ and a regularization constant of $\beta=0.1$. 
Numerical tests reveal that these experiments perform well with 10 re-samplings of the Hessian per iteration and an averaging of 10 Gibbs state preparation per set of Hamiltonian parameters. 
This additional averaging might be necessary due to ill-conditioning of the underlying linear system of equations. In practice, the number of re-samplings and averages can be traded off against a larger standard deviation of the loss function.

Fig.~\ref{fig:generative_learning}(b) shows the development of the loss function of 10 optimization runs of VarQBM with QN-SPSA along with the loss if the exact Gibbs state preparation by means of matrix exponentiation is used. Though they are subject to noise, the VarQBMs reliably converge to the same loss as the optimization with exact evolution. By using more time steps, decreasing the regularization constant $\beta$, and using more Hessian re-samplings, the loss of QN-SPSA, we can attempt to track the exact evolution more closely.

The sampling statistics of the final, trained Gibbs states are presented in Fig.~\ref{fig:generative_learning}(c). The output sampling distribution of the Gibbs state prepared with QN-SPSA approximates the Bell distribution well: the target sampling probabilities for the states $\ket{00}$ and $\ket{11}$ are within the standard deviation of the trained state, and, though they are not exactly 0, the states $\ket{01}$ and $\ket{10}$ have only minuscule amplitudes. These non-zero amplitudes are, however, also present in the final state obtained with exact Gibbs state preparation and might also be a limitation of the chosen system Hamiltonian $\hat{\mathcal{H}}_{\vec\omega}$.

\section{Conclusion}\label{sec:conclusion}

In this paper, we presented how SPSA can be used to approximate the QFIM. Thereby we reduce the cost of circuit executions from scaling quadratically with the number of parameters to constant. We tested the resulting algorithm, QN-SPSA, on ground state preparation and VarQBMs and reproduced existing results from literature where the analytic QFIM was used.

In the ground state calculations, we observed that QN-SPSA inherits the fast convergence and robustness of QNG with respect to the initial parameters, while having the computational cost benefits of SPSA, overall, leading to the most effective optimization method of the tested algorithms. With the reduced number of circuits required to evaluate the QNG, our approach enables the simulation and investigation of much larger systems than previously possible. 
If the system's Hamiltonian is complex and we have to measure in a large number of bases, we have further seen that the overhead of calculating QN-SPSA compared to SPSA becomes negligible. This means we can benefit from the QNG properties at very little additional cost.
There are other proposed optimization routines, designed especially for variational circuits, that minimize the required resources, such as iCANS \cite{kuebler_icans_2020}, and a comparison as well as potential combination with QN-SPSA techniques could be of great interest.

For generative learning with VarQBMs, we successfully trained a Gibbs state to reproduce a Bell-state target distribution.
The speed-up on this small example was not as significant as the ground state calculations due to the required re-sampling. The performance of QN-SPSA for more difficult distributions and more parameters requires further investigation. Another application to consider is VarQBMs for discriminative learning.

QN-SPSA allows incorporating several adaptive strategies to tailor the optimization routine to the problem or improve convergence.
These include using first-order SPSA in the beginning of the optimization to construct a stable QFIM estimate, calibrating the perturbation for the finite difference gradients, and dynamically adjusting the number of sampled dimensions according to the rejected or accepted steps.

A caveat of QN-SPSA is the required evaluation of the overlap of two variational ansatze with different parameters to compute the point-estimates of the QFIM.
Current available algorithms to compute the overlap of two states either require duplication of the circuit depth or of the circuit width, or are not scalable.
Finding options to reduce this overhead is a relevant open question for further research, particularly for running this algorithm on real noisy quantum devices.

To conclude, QN-SPSA provides a promising and efficient new method for parameter optimization in variational quantum algorithms. Given the enormous reduction in the number of evaluations needed for many relevant applications compared with the original QNG, this is an important step towards scaling these quantum algorithms to practically relevant problem sizes.

\section{Acknowledgements}

We would like to thank Ali Javadi-Abhari for insightful discussions throughout the project, in particular regarding the evaluation of state overlaps. We are grateful for Amira Abbas, who generously shared her helpful intuition and knowledge on the Quantum Fisher Information, as well as feedback on this manuscript. Also, we thank Daniel Egger and Panagiotis Barkoutsos for their ideas of challenging loss functions for the optimizers in this work.

Further, we thank Jessie Yu, Andrew Wack, Blake Johnson and the whole Qiskit Runtime team for enabling us to leverage the Qiskit Runtime architecture to significantly improve execution times for the real hardware experiments.

Christa Zoufal acknowledges the support of the National Centre of Competence in Research \emph{Quantum Science and Technology} (QSIT).

We acknowledge the use of IBM Quantum services for this work. The views expressed are those of the authors, and do not reflect the official policy or position of IBM or the IBM Quantum team.

IBM, the IBM logo, and ibm.com are trademarks of International Business Machines Corp., registered in many jurisdictions worldwide. Other product and service names might be trademarks of IBM or other companies. The current list of IBM trademarks is available at \url{https://www.ibm.com/legal/copytrade}.

\bibliographystyle{unsrtnat}
\bibliography{refs}

\onecolumn\newpage
\appendix

\section{Variational Quantum Imaginary Time Evolution}\label{app:varqte}

For a Hamiltonian $\hat{\mathcal{H}}$, the Wick-rotated Schr\"odinger equation 
\begin{equation*}
    \frac{\partial \ket{\psi(t)}}{\partial t} = -(\hat{\mathcal{H}} - E_t)\ket{\psi(t)}
\end{equation*}
describes the normalized form of Quantum Imaginary Time Evolution where
\begin{equation*}
    \ket{\psi(t)} = \frac{e^{-\hat{\mathcal{H}}t}}{\sqrt{\bra{\psi(0)}e^{-\hat{\mathcal{H}}t}\ket{\psi(0)} }}\ket{\psi(0)},
\end{equation*}
and $E_t = \braket{\psi(t)|\hat{\mathcal{H}}|\psi(t)}$.

A time-discretized approximation of this evolution can be implemented using a variational ansatz state $\ket{\psi(\vec\theta^{(t))}}$ with parameters $\vec\theta^{(t)}$ associated with time $t$. VarQITE can be realized using McLachlan's variational principle \cite{mclachlan_variational_1964}
\begin{align*}
    \delta\left\|  \frac{\partial \ket{\psi(\vec\theta^{(t)})}}{\partial t} + (\hat{\mathcal{H}} - E_{\tau})\ket{\psi(\vec\theta^{(t)})} \right\|_{\ell^2}^2 = 0,
\end{align*}
which aims to minimize the distance between the left and right side of the Wick-rotated Schr\"odinger equation w.r.t.~the variational space given by $\ket{\psi(t)}$.

This variational principle leads to the following linear system of equations
\begin{equation*}
    g_{ij}(\vec\theta^{(t)})\frac{\partial\theta_j^{(t)}}{\partial t} = -\text{Re}\left\{\left\langle\frac{\partial\psi(\vec\theta^{(t)})}{\partial \theta_i}\right\vert \hat{\mathcal{H}} \ket{\psi(\vec\theta^{(t)})}\right\}
\end{equation*}
which together with an initial value for $\vec\theta$ defines an initial value problem that can be numerically solved with an ODE solver.

Given that the time discretization used to solve the ODE is chosen sufficiently small and the variational quantum circuit is sufficiently expressive, the VarQITE steps can only decrease the system energy or keep it constant \cite{mcardle_varqite_2019}. 
Thus, this approach offers interesting convergence properties for searching for the state corresponding to the minimum energy eigenstate given that the initial state has a non-zero overlap with this state.

VarQITE may also be used for approximate Gibbs state preparation, see Sec.~\ref{sec:qbm} for further details, as well as for ground state evaluation. The latter case, can be motivated as follows. Suppose the initial state $\ket{\psi(\vec\theta^{(0)})}$ has an overlap with the ground state and the evolution time $t\rightarrow\infty$. In this case, all contributions in $\ket{\psi(\vec\theta^{(t)})}$ which correspond to eigenvalues bigger than the minimum are being damped with time and $\mathrm{lim}_{t \rightarrow \infty} \ket{\psi(\vec\theta^{(t)})}$ is equal to the ground state. Since, in practice, an infinite time cannot be simulated one needs to find a sufficiently big, finite time.

Notably, a VarQITE ground state search coincides with a special case of QNG where
\begin{align*}
    f(\vec\theta) = -\frac{1}{2} \braket{\psi(\vec\theta)|\hat{\mathcal{H}}|\psi(\vec\theta)}.
\end{align*}

\section{Minimization-formulation of Vanilla and Natural Gradient Descent}\label{app:argmin_formulation}

In this section we show the equivalence of Eqs.~\eqref{eq:vanilla_gradient_descent_update} and \eqref{eq:vanilla_argmin} for the vanilla gradient descent update rule by showing that Eq.~\eqref{eq:vanilla_gradient_descent_update} is the solution to the minimization in Eq.~\eqref{eq:vanilla_argmin}.
To solve the minimization 
\begin{equation*}
   \vec\theta^{(k + 1)} = \argmin_{\vec\theta \in \mathbb{R}^d} \left( \langle \vec\theta - \vec\theta^{(k)}, \vec\nabla f(\vec\theta^{(k)}) \rangle + \frac{1}{2\eta} ||\vec\theta - \vec\theta^{(k)}||_2^2\right),
\end{equation*}
we take the gradient of the right hand side with respect to $\vec\theta$ and set it to $\vec 0$
\begin{equation*}
    \vec 0 = \vec\nabla f(\vec\theta^{(k)}) + \frac{1}{\eta} (\vec\theta - \vec\theta^{(k)}).
\end{equation*}
Then, solving for $\vec\theta$ yields the solution labelled $\vec\theta^{(k + 1)}$
\begin{equation*}
    \vec\theta^{(k + 1)} = \vec\theta^{(k)} - \eta \vec\nabla f(\vec\theta^{(k)}),
\end{equation*}
which is exactly Eq.~\eqref{eq:vanilla_gradient_descent_update}.

We solve for the update step of the Natural Gradient Descent
\begin{equation}\notag
   \vec\theta^{(k + 1)} = \argmin_{\vec\theta \in \mathbb{R}^d}\bigg( \langle \vec\theta - \vec\theta^{(k)}, \vec\nabla f(\vec\theta^{(k)}) \rangle + \frac{1}{2\eta} ||\vec\theta - \vec\theta^{(k)}||_{g(\vec\theta^{(k)})}^2\bigg),
\end{equation}
in the same fashion. We differentiate the right-hand side and set it to $\vec 0$
\begin{equation*}
    \vec 0 = \vec\nabla f(\vec\theta^{(k)}) + \frac{1}{\eta} g(\vec\theta^{(k)}) (\vec\theta - \vec\theta^{(k)}),
\end{equation*}
where we used
\begin{equation*}
    \begin{aligned}
    \vec\nabla ||\vec\theta - \vec\theta^{(k)}||^2_{g(\vec\theta^{(k)})} 
    &= \vec\nabla \langle \vec\theta - \vec\theta^{(k)}, g(\vec\theta^{(k)}) (\vec\theta - \vec\theta^{(k)})\rangle \\
    &= g(\vec\theta^{(k)})(\vec\theta - \vec\theta^{(k)})
    + g^T(\vec\theta^{(k)})(\vec\theta - \vec\theta^{(k)}) \\
    &= 2g(\vec\theta^{(k)})(\vec\theta - \vec\theta^{(k)})
    \end{aligned}
\end{equation*}
and the fact that $g$ is symmetric. Then solve for the update step
\begin{equation*}
    \vec\theta^{(k + 1)} = \vec\theta^{(k)} - \eta g^{-1}(\vec\theta^{(k)}) \vec\nabla f(\vec\theta^{(k)}),
\end{equation*}

\section{Comparison of the QFIM Formulas}\label{app:qfi_reformulation}

Eqs.~\eqref{eq:qfi_direct} and~\eqref{eq:qfi_hessian} show different ways to compute the QFI. Here we show the equivalence 
and justify the coefficient of 1/2 in Eq.~\eqref{eq:qfi_hessian}.
A single element of the Fubini-Study metric tensor according to Eq.~\eqref{eq:qfi_hessian} is
\begin{equation*}
    \begin{aligned}
        & -\frac{1}{2} \frac{\partial^2}{\partial\theta_i \partial\theta_j} |\braket{\psi(\vec\theta^\prime) | \psi(\vec\theta)}|^2 \bigg\vert_{\vec\theta^\prime = \vec\theta} \\
        &= -\frac{1}{2} \frac{\partial^2}{\partial\theta_i \partial\theta_j}  \braket{\psi(\vec\theta^\prime) | \psi(\vec\theta)} \braket{\psi(\vec\theta) | \psi(\vec\theta^\prime)} \bigg\vert_{\vec\theta^\prime = \vec\theta} \\
        &= -\frac{\partial}{\partial \theta_i} \mathrm{Re}\left\{\braket{\psi(\vec\theta^\prime) | \psi(\vec\theta)} \left\langle \psi(\vec\theta^\prime) \bigg\vert \frac{\partial \psi}{\partial \theta_j} \right\rangle \right\}\bigg\vert_{\vec\theta^\prime = \vec\theta}\\
        &= -\mathrm{Re}\left\{\left\langle\psi(\vec\theta^\prime) \bigg\vert \frac{\partial^2\psi}{\partial\theta_i\partial\theta_j} \right\rangle + \left\langle\psi(\vec\theta^\prime) \bigg\vert \frac{\partial \psi}{\partial \theta_j}\right\rangle \left\langle\frac{\partial \psi}{\partial \theta_i} \bigg\vert \psi(\vec\theta^\prime) \right\rangle\right\} \bigg\vert_{\vec\theta^\prime = \vec\theta}\\
        &= \mathrm{Re}\left\{-\left\langle\psi(\vec\theta) \bigg\vert \frac{\partial^2\psi}{\partial \theta_i\partial \theta_j}\right\rangle - \left\langle\frac{\partial \psi}{\partial \theta_i} \bigg\vert\psi(\vec\theta) \right\rangle \left\langle\psi(\vec\theta) \bigg\vert  \frac{\partial\psi}{\partial \theta_j} \right\rangle\right\}.
    \end{aligned}
\end{equation*}
We can rewrite the first summand using the identity we obtain from differentiating both sides of the equation $1 = \braket{\psi(\vec\theta)|\psi(\vec\theta)}$ with 
respect to $\vec\theta$,
\begin{equation*}
    \begin{aligned}
    0 &= \frac{\partial^2}{\partial \theta_i \partial \theta_j} \braket{\psi(\vec\theta)|\psi(\vec\theta)} \\
      &= 2 \mathrm{Re}\left\{\frac{\partial}{\partial \theta_i} \left\langle\frac{\partial\psi}{\partial \theta_j}\bigg\vert\psi(\vec\theta)\right\rangle\right\}  \\
      &= 2 \mathrm{Re}\left\{\left(\left\langle\frac{\partial^2\psi}{\partial \theta_i \partial \theta_j}\bigg\vert\psi(\vec\theta)\right\rangle + \left\langle\frac{\partial\psi}{\partial \theta_i}\bigg\vert\frac{\partial\psi}{\partial \theta_j}\right\rangle\right) \right\}\\
   \Leftrightarrow& -\mathrm{Re}\left\{\left\langle\frac{\partial^2\psi}{\partial \theta_i\partial \theta_j}\bigg\vert\psi(\vec\theta)\right\rangle\right\} = \mathrm{Re}\left\{\left\langle\frac{\partial\psi}{\partial \theta_i}\bigg\vert\frac{\partial\psi}{\partial \theta_j}\right\rangle\right\}.
    \end{aligned}
\end{equation*}
Replacing the second derivative of $\ket{\psi(\vec\theta)}$ with the two first order derivatives we obtain
\begin{equation*}
    g_{ij}(\vec\theta) = \mathrm{Re}\left\{\left\langle\frac{\partial \psi}{\partial \theta_i} \bigg\vert \frac{\partial \psi}{\partial \theta_j}\right\rangle - \left\langle\frac{\partial \psi}{\partial \theta_i} \bigg\vert  \psi(\vec\theta) \right\rangle \left\langle\psi(\vec\theta) \bigg\vert \frac{\partial\psi}{\partial \theta_j} \right\rangle\right\},
\end{equation*}
which is the same as Eq.~\eqref{eq:qfi_direct}.

\section{Influence of QFIM Regularization}\label{app:qfi_regularization}

In each iteration step QN-SPSA constructs a (up to) rank-2 estimate $\hat g^{(k)}$ of the QFI.
Even though we start with a full rank matrix, $\bar g^{(0)} = I$ this can lead to a singular 
approximation.
Since we have to invert the estimated QFI, or solve a LSE, this can be problematic.

To solve this problem we ensure that the estimate is symmetric-positive definite in each iteration by taking the absolute value of the eigenvalues and adding a regularization constant $\beta > 0$ on the diagonal, i.e., $\sqrt{A A} + \beta I$.
To better understand the influence of the regularization for larger values we further normalize the
expression as $(\sqrt{A A} + \beta I) / (1 + \beta)$. 
Without the normalization, a large regularization
decreases the step size such that for $\beta \rightarrow \infty$ the update step approaches zero.

If the QFIM estimate is faithful the eigenvalues are already positive and taking the absolute value of the eigenvalues 
does not change anything.
However, adding the regularization constant always has an impact. 
A small regularization constant leads to an update closer to the QNG while a large constant neglects the QFIM approximation and leads to an update closer to vanilla gradient descent.
On the other hand, using a small constant is more prone to numerical instabilities than using a large one.

\begin{figure}
    \centering
    \includegraphics[width=0.4\linewidth]{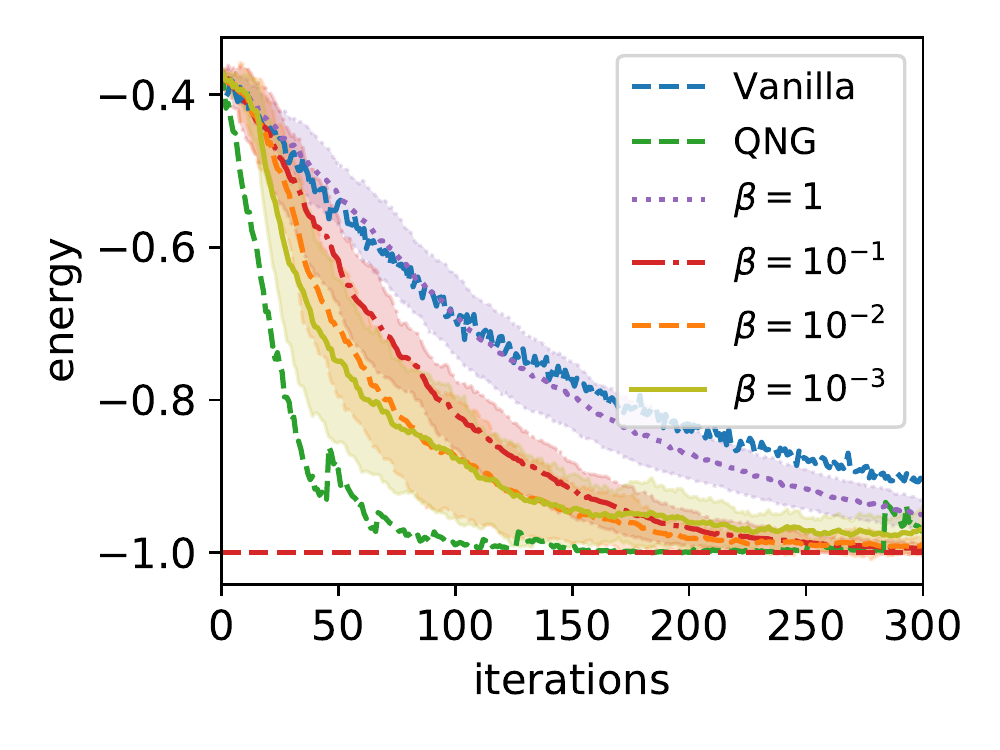}
    \caption{Loss function for the Pauli two-design for vanilla and natural gradient descent and QN-SPSA. QN-SPSA is shown for different regularization constants $\beta \in \{10^{-3}, 10^{-2}, 10^{-1}, 1\}$.}
    \label{fig:regularization}
\end{figure}

The regularization $\beta$ is thus a hyper-parameter to trade off numerical stability for faithful QFIM approximation. Fig.~\ref{fig:regularization} shows the Pauli two-design example from 
Sec.~\ref{sec:numerics_qng} with 9 qubits for different values of $\beta$ and visualizes how the regularization can be used to interpolate between natural and vanilla gradient descent.

\section{Convergence efficiency of natural gradients}\label{app:maxcut}

In Secs.~\ref{sec:numerics_qng} and~\ref{sec:lih} we show how fast different gradient descent techniques converge in the number of required iteration steps and in the number of required circuit evaluations. We observed, that natural gradients have a converge clearly faster than vanilla gradient approaches in the number of iterations, however due to the additional costs, they might not be more efficient if the circuit costs are considered. Due to the small overhead of QN-SPSA, especially if the system Hamiltonian requires a lot of measurements. Here, we investigate another example where QN-SPSA is more efficient than first-order SPSA even if the Hamiltonian is diagonal and require only one measurement per loss function evaluation.

In difficult loss landscapes, where small changes in a subset of parameters can lead to large changes in function values, natural gradient methods show a stable convergence since they control how much a parameter step can change the model. Vanilla gradients are agnostic to model sensitivity and easily overshoot if the learning rate is not adjusted to the loss function.

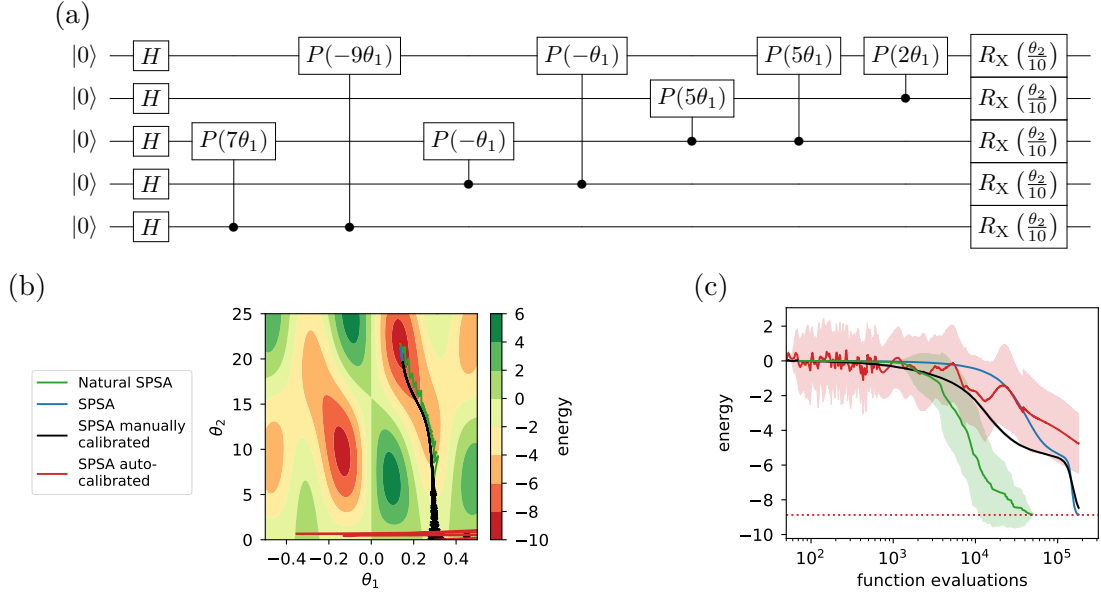
\begin{figure*}
    \centering
    \begin{tikzpicture}
        \node[anchor=north west] at (-0.5, 0.5) {\includegraphics[width=0.8\textwidth]{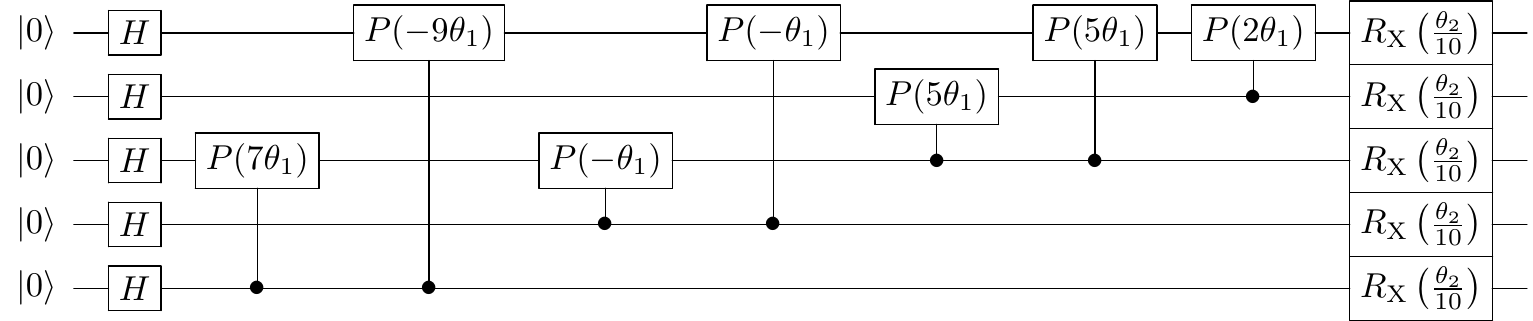}};
        \node[anchor=north west] at (-1, -3) {\includegraphics[height=4cm]{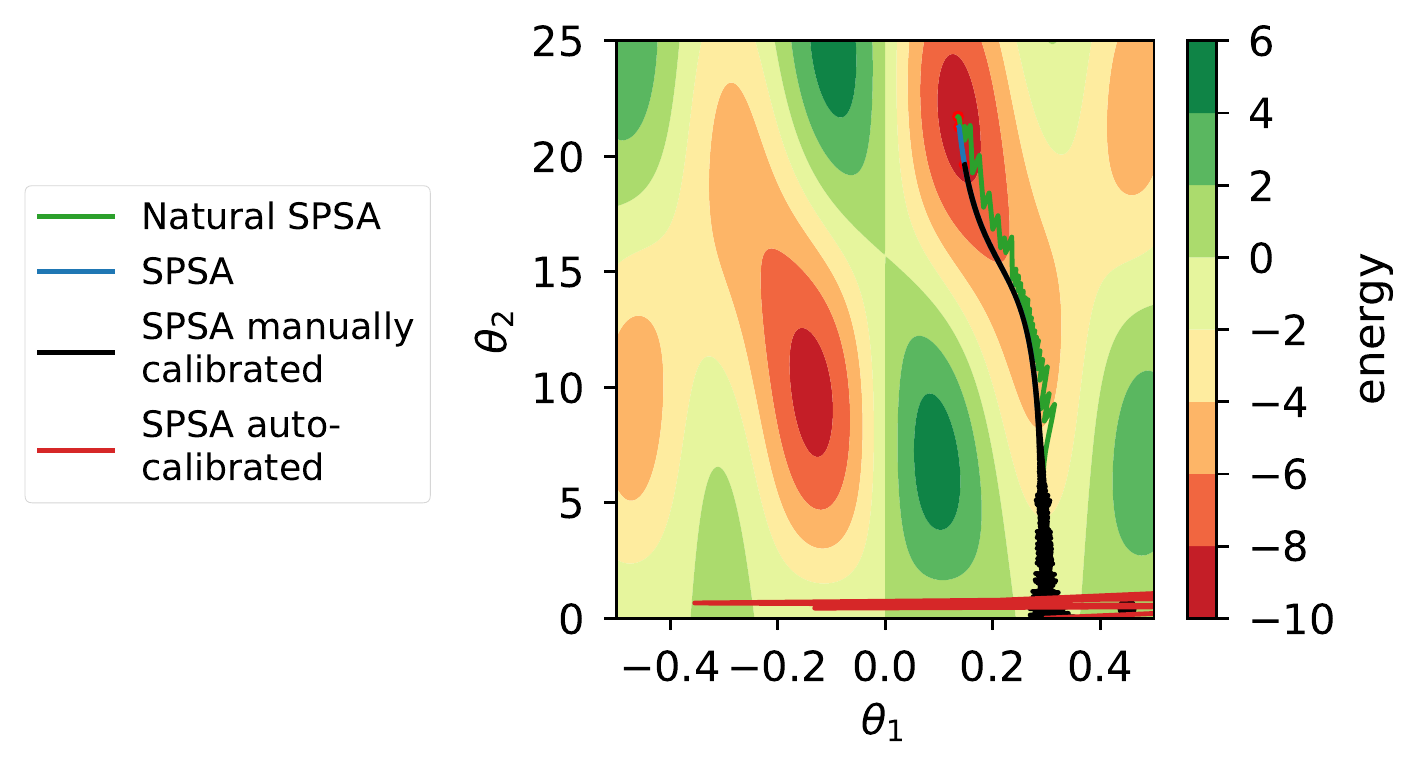}};
        \node[anchor=north west] at (8, -3) {\includegraphics[height=4cm]{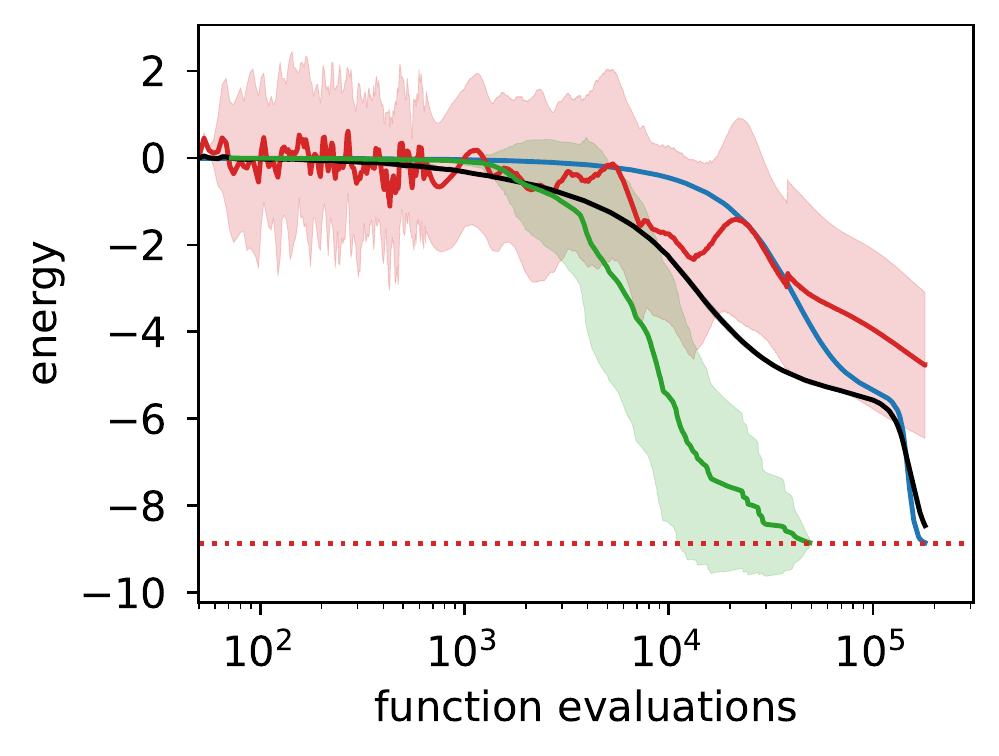}};
        \node at (-0.2, 0.6) {(a)};
        \node at (-0.8, -3) {(b)};
        \node at (8.2, -3) {(c)};
    \end{tikzpicture}
    \caption{The weighted MAXCUT problem. (a) The QAOA ansatz for the problem instance. (b) The loss landscape and the path of different optimization routines through the landscape. (c) The convergence of the investigates methods with respect to the required number of function evaluations.}
    \label{fig:maxcut}
\end{figure*}

To investigate this behaviour, we consider a challenging loss landscape motivated
from a QAOA ansatz \cite{farhi_qaoa_2014} for a MAXCUT problem on a five-node random graph with random integer weights sampled from $\mathcal{U}([-10, 10])$. 
The observable for this particular application is
\begin{equation*}
    \begin{aligned}
   \hat{\mathcal{H}} =& Z_4 Z_5 + 2.5 Z_3 Z_5 + 2.5 Z_3 Z_4 - 0.5 Z_2 Z_5 \\
          &- 0.5 Z_2 Z_3 - 4.5 Z_1 Z_5 + 3.5 Z_1 Z_3
    \end{aligned}
\end{equation*}
and the mixer Hamiltonian is $\sum_{i=1}^{5} X_i / 20$.
The resulting QAOA ansatz is shown in Fig.~\ref{fig:maxcut}(a).

The resulting loss function has numerous local extrema and has spiky regions close to the global
minima, where the gradients are several magnitudes larger than in the flatter surroundings. 
The paths of the different optimizers in this landscape is shown in Fig.~\ref{fig:maxcut}(b). 
QN-SPSA uses a learning rate of $\eta=10^{-2}$, a displacement of $\epsilon=10^{-2}$ and a regularization constant 
of $\beta = 10^{-3}$.
SPSA is run three times: first with the same settings as QN-SPSA, then with the automatic calibration introduced in \cite{kandala_hardware_2017}, and lastly with a manually adjusted calibration.
The automated calibration chooses the learning rate and displacements as power series with optimal exponents for SPSA \cite{spall_spsa_1998} and calibrates the constant coefficients of the power series such that in the first step the magnitude of the parameter update is $|\theta^{(1)}_i - \theta^{(0)}_i| \approx 2\pi/10$.
However, in practice, fixing the parameter update can be problematic as it does not take into account how sensitive the model is with respect to a rescaling of the parameters. This becomes obvious in Fig.~\ref{fig:maxcut}(b), where SPSA with this automatic calibration acts on too large length scales and starts to oscillate heavily.
Thus, in the second run, we manually tested different parameter magnitude updates and selected the best at $|\theta^{(1)}_i - \theta^{(0)}_i| \approx 0.1$. 

In Fig.~\ref{fig:maxcut}(c), the mean and standard deviation of the loss for 25 runs is shown for each of the methods.
We clearly see that QN-SPSA outperforms SPSA, even with manual calibrations and the additional evaluation costs taken into account.

The MAXCUT experiment highlights another advantage of natural gradient approaches. Vanilla gradient descent optimizations, both analytic and SPSA-based, require careful tuning of the learning rate to the sensitivity of the objective function. An optimal tuning might further not always be possible since the learning rate acts globally in each parameter dimension but the objective might necessarily be equally sensitive in each dimension. In natural gradient methods, the learning rate controls the change of the model instead of the parameters and takes the objective sensitivity into account. Thus, the learning rate can largely be set independent of the model and in practice the value of $\eta=10^{-2}$ worked well on every example.

\end{document}